\PassOptionsToPackage{table}{xcolor}
\documentclass[sigconf]{acmart}

\AtBeginDocument{%
  }

\copyrightyear{2026}
\acmYear{2026}
\setcopyright{cc}
\setcctype{by}
\acmConference[SoCC '26]{ACM Symposium on Cloud Computing V.2}{November 18--20, 2026}{Singapore, Singapore}
\acmBooktitle{ACM Symposium on Cloud Computing V.2 (SoCC '26), November 18--20, 2026, Singapore, Singapore}
\acmDOI{10.1145/3828161.3857850}
\acmISBN{979-8-4007-2816-7/2026/11}

\newcommand{\secref}[1]{\S\ref{#1}}
\usepackage{xcolor}

\makeatletter
 
\newcommand\affiliationsep{\quad\ }
\newcommand\affiliationfieldsep{, }
 
\ExplSyntaxOn
 
\int_new:N  \g__ia_author_int
\int_new:N  \g__ia_aff_int
\seq_new:N  \g__ia_group_seq
\bool_new:N \g__ia_group_closed_bool
\prop_new:N \g__ia_aff_prop
\clist_new:N \g__ia_breaks_clist
\clist_new:N \g__ia_affbreaks_clist
\tl_new:N   \l__ia_num_tl
\bool_new:N \l__ia_first_bool
\seq_new:N  \g__ia_inst_seq
\seq_new:N  \g__ia_dept_seq
 
\prg_generate_conditional_variant:Nnn \seq_if_in:Nn { cV } { F }
 
\NewDocumentCommand \breakauthorsafter { m }
  { \clist_gset:Nn \g__ia_breaks_clist {#1} }
\NewDocumentCommand \breakaffiliationsafter { m }
  { \clist_gset:Nn \g__ia_affbreaks_clist {#1} }
 
\AtBeginMaketitle
  {
    \if@ACM@anonymous \else
      \int_compare:nNnT { \value{footnote} } < { 1 } { \setcounter{footnote}{1} }
      \tl_gput_left:Nn \@titlenotes
        { \int_compare:nNnT { \value{footnote} } < { 1 } { \setcounter{footnote}{1} } }
    \fi
  }
 
\cs_new_protected:Npn \__ia_new_author:n #1
  {
    \bool_if:NT \g__ia_group_closed_bool
      {
        \seq_gclear:N \g__ia_group_seq
        \bool_gset_false:N \g__ia_group_closed_bool
      }
    \int_gincr:N \g__ia_author_int
    \tl_gclear_new:c  { g__ia_name_  \int_use:N \g__ia_author_int _tl }
    \tl_gset:cn       { g__ia_name_  \int_use:N \g__ia_author_int _tl } {#1}
    \seq_gclear_new:c { g__ia_affs_  \int_use:N \g__ia_author_int _seq }
    \seq_gclear_new:c { g__ia_marks_ \int_use:N \g__ia_author_int _seq }
    \seq_gput_right:Nx \g__ia_group_seq { \int_use:N \g__ia_author_int }
  }
 
\cs_new_protected:Npn \__ia_add_aff:n #1
  {
    \bool_gset_true:N \g__ia_group_closed_bool
    \prop_get:NnNF \g__ia_aff_prop {#1} \l__ia_num_tl
      {
        \int_gincr:N \g__ia_aff_int
        \tl_set:Nx \l__ia_num_tl { \int_use:N \g__ia_aff_int }
        \prop_gput:NnV \g__ia_aff_prop {#1} \l__ia_num_tl
        \tl_gclear_new:c { g__ia_afftext_ \l__ia_num_tl _tl }
        \tl_gset:cn      { g__ia_afftext_ \l__ia_num_tl _tl } {#1}
      }
    \seq_map_inline:Nn \g__ia_group_seq
      {
        \seq_if_in:cVF { g__ia_affs_ ##1 _seq } \l__ia_num_tl
          { \seq_gput_right:cV { g__ia_affs_ ##1 _seq } \l__ia_num_tl }
      }
  }
 
\cs_new_protected:Npn \__ia_add_mark:n #1
  {
    \int_compare:nNnT { \g__ia_author_int } > { 0 }
      { \seq_gput_right:cn { g__ia_marks_ \int_use:N \g__ia_author_int _seq } {#1} }
  }
 
\cs_new_protected:Npn \__ia_supers:n #1
  {
    \bool_lazy_and:nnF
      { \seq_if_empty_p:c { g__ia_affs_  #1 _seq } }
      { \seq_if_empty_p:c { g__ia_marks_ #1 _seq } }
      {
        \textsuperscript
          {
            \seq_use:cn { g__ia_affs_ #1 _seq } { , }
            \bool_set:Nn \l__ia_first_bool
              { \seq_if_empty_p:c { g__ia_affs_ #1 _seq } }
            \seq_map_inline:cn { g__ia_marks_ #1 _seq }
              {
                \bool_if:NTF \l__ia_first_bool
                  { \bool_set_false:N \l__ia_first_bool }
                  { , }
                ##1
              }
          }
      }
  }
 
\cs_new_protected:Npn \__ia_names:
  {
    \int_step_inline:nn { \g__ia_author_int }
      {
        \mbox { \tl_use:c { g__ia_name_ ##1 _tl } \__ia_supers:n {##1} }
        \int_compare:nNnT {##1} < { \g__ia_author_int }
          {
            ,
            \clist_if_in:NnTF \g__ia_breaks_clist {##1} { \\ } { ~ }
          }
      }
  }
 
\cs_new_protected:Npn \__ia_aff_format:n #1
  {
    \group_begin:
      \seq_gclear:N \g__ia_inst_seq
      \seq_gclear:N \g__ia_dept_seq
      \cs_set_protected:Npn \__ia_skip:n ##1 { }
      \cs_set_protected:Npn \institution ##1 { \seq_gput_right:Nn \g__ia_inst_seq {##1} }
      \RenewDocumentCommand \department { O{} m } { \seq_gput_right:Nn \g__ia_dept_seq {##2} }
      \cs_set_eq:NN \city          \__ia_skip:n
      \cs_set_eq:NN \state         \__ia_skip:n
      \cs_set_eq:NN \country       \__ia_skip:n
      \cs_set_eq:NN \position      \__ia_skip:n
      \cs_set_eq:NN \streetaddress \__ia_skip:n
      \cs_set_eq:NN \postcode      \__ia_skip:n
      \hbox_set:Nn \l_tmpa_box {#1}
      \seq_concat:NNN \l_tmpa_seq \g__ia_inst_seq \g__ia_dept_seq
      \seq_use:Nn \l_tmpa_seq { \affiliationfieldsep }
    \group_end:
  }
\cs_generate_variant:Nn \__ia_aff_format:n { v }
 
\cs_new_protected:Npn \__ia_affs:
  {
    \int_step_inline:nn { \g__ia_aff_int }
      {
        \mbox
          {
            \textsuperscript{##1}
            \__ia_aff_format:v { g__ia_afftext_ ##1 _tl }
          }
        \int_compare:nNnT {##1} < { \g__ia_aff_int }
          {
            \clist_if_in:NnTF \g__ia_affbreaks_clist {##1}
              { \\ } { \affiliationsep }
          }
      }
  }
 
\cs_new_protected:Npn \__ia_mkauthors:
  {
    \cs_gset:Npn \@currentauthors { }
    \cs_gset:Npn \@currentaffiliation { }
    \group_begin:
      \dim_set:Nn \hsize { \textwidth }
      \cs_set_protected:Npn \@author ##1 { \__ia_new_author:n {##1} }
      \cs_set_protected:Npn \and { }
      \cs_set_protected:Npn \affiliation ##1 ##2 { \__ia_add_aff:n {##2} }
      \cs_set_protected:Npn \email ##1 ##2 { }
      \cs_set_protected:Npn \@authornotemark
        { \__ia_add_mark:n { \stepcounter{footnote} \thefootnote } }
      \cs_set_protected:Npn \@@authornotemark ##1
        {
          \__ia_add_mark:n
            { \group_begin: \c@footnote = ##1 \scan_stop: \thefootnote \group_end: }
        }
      \cs_set_protected:Npn \@correspondingauthormark
        { \__ia_add_mark:n { * } }
      \addresses
      \global\setbox\mktitle@bx = \vbox
        {
          \noindent \unvbox\mktitle@bx \par \medskip
          \group_begin:
            \centering \@authorfont \__ia_names: \par
          \group_end:
          \int_compare:nNnT { \g__ia_aff_int } > { 0 }
            {
              \group_begin:
                \centering \@affiliationfont \__ia_affs: \par
              \group_end:
            }
          \bigskip
        }
    \group_end:
  }
 
\cs_new_eq:NN \__ia_orig_mkauthors: \@mkauthors
\cs_gset_protected:Npn \@mkauthors
  {
    \if@ACM@anonymous
      \__ia_orig_mkauthors:
    \else
      \__ia_mkauthors:
    \fi
  }
 
\ExplSyntaxOff
\makeatother
\breakauthorsafter{3}
\breakaffiliationsafter{2}

\usepackage{xspace}
\usepackage{enumitem}
\newcommand{\dynsl}{\textsc{DynSL}\xspace}

\usepackage[capitalise]{cleveref}

\usepackage{pifont}% http://ctan.org/pkg/pifont
\newcommand{\cmark}{\ding{51}}%
\newcommand{\xmark}{\ding{55}}%

\usepackage{multirow}
\usepackage{algorithm}
\usepackage[noend]{algorithmic}
\newtheorem{claim}{Claim}

\usepackage{balance}
\definecolor{gold}{HTML}{FBF2D2}
\definecolor{silver}{HTML}{DDDDDD}
\definecolor{bronze}{HTML}{EED2B8}
\definecolor{goldD}{HTML}{D9AE13}
\definecolor{silverD}{HTML}{909090}
\definecolor{bronzeD}{HTML}{9A5F26}
\usepackage{balance}
\usepackage{calc}
\newcommand{\medalsample}{000.0}
\newlength{\medalwd}
\newcommand{\medal}[3]{{%
  \setlength{\medalwd}{\maxof{\widthof{\medalsample}}{\widthof{#3}}}%
  \setlength{\fboxsep}{1.5pt}%
  \fcolorbox{#2}{#1}{\makebox[\medalwd]{#3}}}}

\newcommand{\medalrank}[2]{%
    \ifcase#1 #2% case 0: no medal
    \or % case 1
      \medal{gold}{goldD}{#2}%
    \or % case 2
      \medal{silver}{silverD}{#2}%
    \or % case 3
      \medal{bronze}{bronzeD}{#2}%
    \else % default case
      #2%
    \fi\ignorespaces
}

\newlength{\meanwd}
\newlength{\stdwd}
\newcommand{\meanstd}[3][0]{{%
  \setlength{\meanwd}{\widthof{\medal{gold}{goldD}{\medalsample}}}%
  \setlength{\stdwd}{\widthof{\tiny{ $\pm$ 0.0}}}%
  \makebox[\meanwd]{\medalrank{#1}{#2}}%
  \makebox[\stdwd][l]{\tiny{ $\pm$ #3}}}}

\newif\ifsubmission
\submissionfalse
\submissiontrue

\ifsubmission
\newcommand{\borisnote}[1]{}
\newcommand{\mcnote}[1]{}
\newcommand{\iea}[1]{}
\newcommand{\rc}[1]{}
\newcommand{\borisinline}[1]{}
\else
\usepackage[textsize=small]{todonotes}
\newcommand{\borisnote}[1]{\todo[color=cyan,inline]{boris: #1}}
\newcommand{\mcnote}[1]{\todo[color=purple!40,inline]{MC: #1}}
\newcommand{\borisinline}[1]{\textit{\textcolor{orange}{[boris]: #1}}}
\newcommand{\iea}[1]{\textit{\textcolor{red}{[iea]: #1}}}
\newcommand{\rc}[1]{\textit{\textcolor{blue}{[rc]: #1}}}
\fi

\AtBeginDocument{%
  \setlength{\abovedisplayskip}{2pt plus 1pt minus 1pt}%
  \setlength{\belowdisplayskip}{2pt plus 1pt minus 1pt}%
  \setlength{\abovedisplayshortskip}{0pt plus 2pt}%
  \setlength{\belowdisplayshortskip}{2pt plus 2pt minus 1pt}%
}

\newcommand{\replaced}[2]{#2}
\newdimen\ulgap   \ulgap=1.0pt   % baseline-to-rule distance (smaller = higher)
\newdimen\ulth    \ulth=0.4pt    % rule thickness
\def\ulhalo{1.5}                 % gap around descenders, in bp

\ifdefined\pdfextension
  \protected\def\ulpdf#1{\pdfextension literal page{#1}}   % LuaLaTeX
\else\ifdefined\pdfliteral
  \def\ulpdf#1{\pdfliteral page{#1}}                       % pdfLaTeX
\else
  \def\ulpdf#1{}            % XeLaTeX
\fi\fi

\newcommand\ulx[1]{%
  \leavevmode
  \setbox0=\hbox{#1}%
  \rlap{\vrule height -\ulgap depth \dimexpr\ulgap+\ulth\relax width \wd0}%
  \rlap{\ulpdf{q 1 g 1 G 2 Tr \ulhalo\space w 1 j}\copy0\ulpdf{Q}}%
  \box0}

\makeatletter
\def\ul@space{\leaders\hrule height -\ulgap depth \dimexpr\ulgap+\ulth\relax
  \hskip\fontdimen2\font plus\fontdimen3\font minus\fontdimen4\font\relax}

\def\ul@words#1 #2\@nil{%
  \ulx{#1}%
  \ifx\relax#2\relax\else\ul@space\ul@words#2\@nil\fi}

\newcommand\ul[1]{\ul@words#1 \@nil}
\makeatother
\newtheorem{assumption}{Assumption}
\newcommand{\statement}[4]{% #1 counter, #2 head, #3 label, #4 text
  \par\smallskip
  \leavevmode\refstepcounter{#1}\label{#3}%
  \textsc{#2~\arabic{#1}.}\enspace\textit{#4}%
  \par\smallskip}
\newcommand{\Observation}[2]{\statement{observation}{Observation}{#1}{#2}}

\usepackage{subcaption}

\makeatletter
\newcommand{\borisparagraph}[1]{%
  \par\textbf{#1}\space
}
\makeatother

\begin{document}

\title[DynSL: Resource-Adaptive Model Training across Heterogeneous Edge Devices]{DynSL: Resource-Adaptive Model Training across\\Heterogeneous Edge Devices}

% \affiliation{%
% \author{Boris Radovič\textsuperscript{1,2,\dag},
%         Istemi Ekin Akkus\textsuperscript{3},
%         Marco Canini\textsuperscript{1},\\
%         Ruichuan Chen\textsuperscript{3},
%         Alice Dethise\textsuperscript{3},
%         Ivica Rimac\textsuperscript{3},
%         Veljko Pejović\textsuperscript{2,4}
% }
% \affiliation[obeypunctuation=true]{%
% \institution{
%     \textsuperscript{1}KAUST \quad
%      \textsuperscript{2}University of Ljubljana, Faculty of Computer and Information Science\quad   
%     \\
%     \textsuperscript{3}Nokia Bell Labs \quad
%     \textsuperscript{4}Jožef Stefan Institute, Department of Computer Systems
% }%
% \city{}\country{}
% }
\author{Boris Radovič}
\authornote{Work partly done while the author was an intern at Nokia Bell Labs.}   % the † footnote
\affiliation{\institution{KAUST}\city{Thuwal}\country{Saudi Arabia}}
\affiliation{\institution{University of Ljubljana}
  \department{Faculty of Computer and Information Science}
  \city{Ljubljana}\country{Slovenia}}

\author{Istemi Ekin Akkus}
\affiliation{\institution{Nokia Bell Labs}\city{Stuttgart}\country{Germany}}

\author{Marco Canini}
\affiliation{\institution{KAUST}\city{Thuwal}\country{Saudi Arabia}}

\author{Ruichuan Chen}
\author{Alice Dethise}
\author{Ivica Rimac}
\affiliation{\institution{Nokia Bell Labs}\city{Stuttgart}\country{Germany}}

\author{Veljko Pejović}
\affiliation{\institution{University of Ljubljana}
  \department{Faculty of Computer and Information Science}
  \city{Ljubljana}\country{Slovenia}}
\affiliation{\institution{Jožef Stefan Institute}
  \department{Department of Computer Systems}
  \city{Ljubljana}\country{Slovenia}}

\renewcommand{\authors}{Boris Radovič, Istemi Ekin Akkus, Marco Canini, Ruichuan Chen, Alice Dethise, Ivica Rimac, and Veljko Pejović}

% \thanks{\textsuperscript{\dag} Work partly done while the author was an intern at Nokia Bell Labs.}

% % Toggle: \crtrue => show only the new text, \crfalse => show red/green markup
% \newtoggle{cr}
% \toggletrue{cr}      % camera-ready: only "text after"
% % \togglefalse{cr}    % revision mode: red struck-through + green

% \newcommand{\replaced}[2]{%
%   \iftoggle{cr}%
%     {#2}%
%     {\textcolor{red}{\sout{#1}}\,\textcolor{green!50!black}{#2}}%
% }

\renewcommand{\shortauthors}{Radovič et al.}
%%
%% Article type: Research, Review, Discussion, Invited or position
% \acmArticleType{Research}
%%
%% Links to code and data
% \acmCodeLink{https://github.com/borisveytsman/acmart}
% \acmDataLink{htps://zenodo.org/link}
%%
%% Authors' contribution
% \acmContributions{BT and GKMT designed the study; LT, VB, and AP
%   conducted the experiments, BR, HC, CP and JS analyzed the results,
%   JPK developed analytical predictions, all authors participated in
%   writing the manuscript.}
%%
%% Sometimes the addresses are too long to fit on the page.  In this
%% case uncomment the lines below and fill them accodingly.
%%
%% \authorsaddresses{Corresponding author: Ben Trovato,
%% \href{mailto:trovato@corporation.com}{trovato@corporation.com};
%% Institute for Clarity in Documentation, P.O. Box 1212, Dublin,
%% Ohio, USA, 43017-6221}
\begin{abstract}
Training\replaced{ machine learning}{} models over data distributed across edge devices presents a\replaced{ fundamental}{} trade-off: centralizing data simplifies training but raises privacy and regulatory concerns, while full local training may exceed the\replaced{}{ devices' }compute and memory capacity\replaced{ of edge devices}{}. Split learning addresses this challenge by partitioning the model between clients and the server, allowing part of the computation to be offloaded. However, existing approaches typically rely on static split configurations or simplified assumptions that fail to capture the heterogeneity and complex \replaced{system }{}dynamics of real deployments. We present \dynsl, a framework that dynamically allocates model layers between clients and the server\replaced{ to minimize training time. \dynsl}{ by} continuously \replaced{monitors}{monitoring} system metrics and periodically \replaced{solves}{solving} an optimization problem that accounts for client and server computation \replaced{as well as}{and} communication\replaced{ costs}{}. When beneficial, the system reconfigures the split\replaced{ during training}{}, enabling the offloading strategy to adapt to changing device capabilities and system conditions. Extensive experiments on a \replaced{real-world }{}testbed of 24 heterogeneous edge devices show that \dynsl reduces training time by up to $\sim\!30\%$ over the strongest adaptive baseline and by more than $50\%$ over a\replaced{n uninformed fixed}{ suboptimal uniform} split, with no loss -- and sometimes a gain -- in \replaced{final }{}accuracy.
\end{abstract}


\authorsaddresses{}
\maketitle

\section{Introduction}

As the volume of data generated at the network edge --- such as on smartphones, IoT sensors, and wearable devices --- continues to grow, industry and academia seek efficient mechanisms to leverage such data for training machine learning models. Federated Learning (FL)~\cite{McMahan2017-qj,Peter_Kairouz_etal2021-bv} has emerged as a prominent paradigm\replaced{ for this purpose. By}{: by} enabling collaborative training across a population of mobile/edge devices without requiring data centralization, \replaced{FL}{it} allows models to benefit from distributed data while preserving user privacy.

\replaced{Despite its conceptual appeal, practical deployments of FL face substantial challenges due to the inherent heterogeneity of edge devices~\cite{Bonawitz2019-zg}. In standard FL, each participating client is required to train the full model on its local infrastructure. However, clients may differ significantly in dataset sizes, computational capabilities, memory availability, network bandwidth, background workload, and more generally in hardware and software configurations~\cite{Bovzivc2024-ms}. As a result, when training a model locally, times can vary considerably across clients, with some requiring significantly more time than their peers. Such clients are commonly referred to as \emph{stragglers}, as they slow down the overall training process~\cite{li2020federated}.}{}

\replaced{In more extreme cases, resource-constrained clients may even be unable to participate in FL if the model is too large to train within acceptable time or memory limits~\cite{niu2023overcoming}. Excluding such clients can bias the trained model toward the data distribution of more powerful devices, potentially harming both model fairness and generalization~\cite{benmalek2024bias,mukhtiar2025fairness,mao2024federated}. To mitigate such issues, some approaches train only sub-parts of the global model on resource-limited clients~\cite{Diao2020-ge,Horvath2021-aa,Yuan2022-ms,Caldas2018-ki}, while others train models tailored to client resources and transfer knowledge via logits rather than model parameters~\cite{He2020-cu,Lin2020-ug,Li2019-qc}. Although effective in certain settings, these techniques often lead to substandard predictive performance~\cite{Radovic2024-do}.}{}

\replaced{}{A key obstacle to deploying FL is the heterogeneity of edge devices~\cite{Bonawitz2019-zg}, which differ in\replaced{ dataset size,}{} compute, memory, network\replaced{ bandwidth}{}, background workload, and other\replaced{ hardware and software}{} factors~\cite{Bovzivc2024-ms}. Since each client must train the full model locally, training times vary considerably, and the slowest clients -- \emph{stragglers} -- delay the whole training process~\cite{li2020federated}. Clients whose resources cannot accommodate the model may even be excluded~\cite{niu2023overcoming}, biasing the model toward the data of more powerful devices and harming fairness and generalization~\cite{benmalek2024bias,mukhtiar2025fairness,mao2024federated}. To mitigate such issues, some approaches train only sub-parts of the global model on resource-limited clients~\cite{Diao2020-ge,Horvath2021-aa,Yuan2022-ms,Caldas2018-ki}, while others train models tailored to client resources~\cite{He2020-cu,Lin2020-ug,Li2019-qc}. Although effective in certain settings, these techniques often lead to lower accuracy~\cite{Radovic2024-do}.}

\begin{figure}[tbp]
    \centering
    \includegraphics[width=0.98\linewidth]{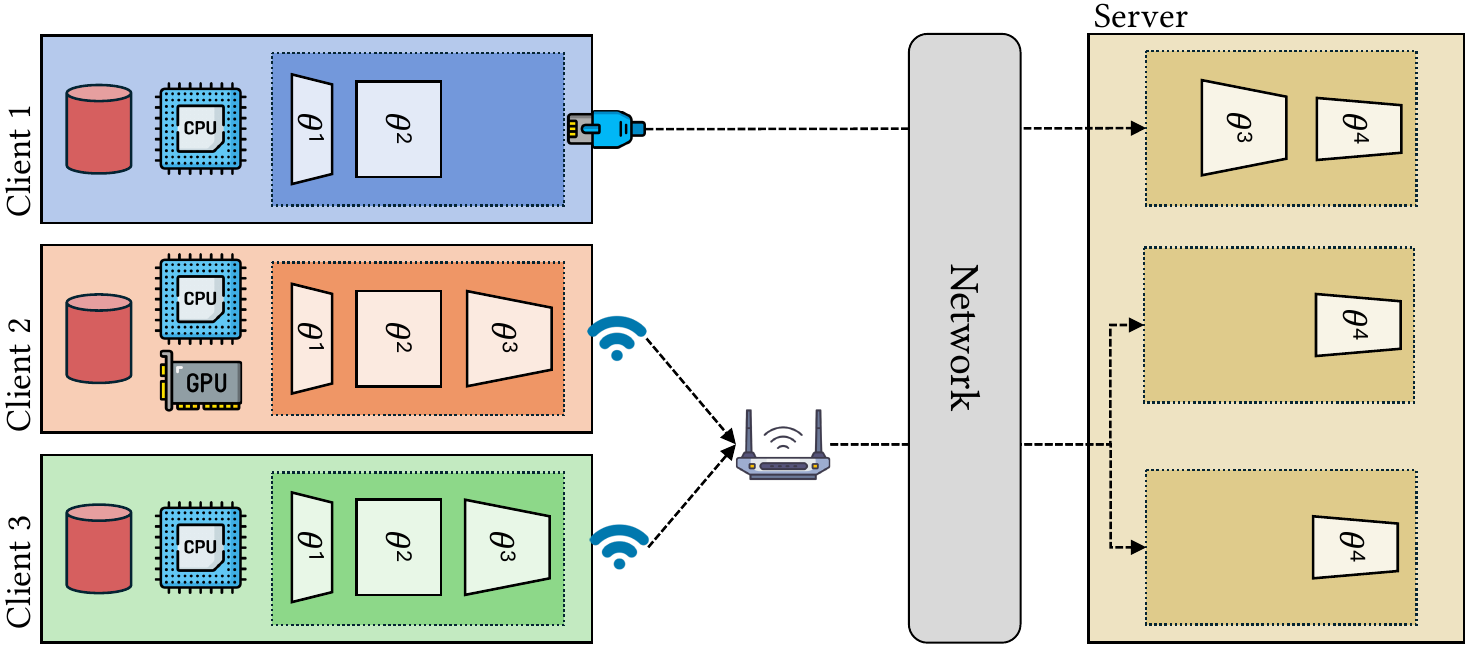}
    % \caption{Example of computation offloading. Each node trains the layers it is assigned, with embeddings and gradients being exchanged to make training possible. Client 1 has no GPU, thus processes $2$ layers. Client 2 has a GPU so it trains $3$ layers. Client 3, while not having a GPU, processes $3$ layers to decrease the embedding size as it communicates through slow WiFi.}
    \caption{\replaced{In FSL, the optimal split depends jointly on compute and network. Client $2$ trains more layers because it has a GPU. Client $3$ (no GPU, slow WiFi) also trains more layers --- not for compute, but to shrink the embeddings it must transmit --- while Client $1$ (no GPU, fast link) trains fewer layers.}{The optimal split depends on both compute and network. Client 2 trains more layers as it has a GPU; Client 3 (no GPU, slow WiFi) does too, but to exchange smaller embeddings; Client 1 (no GPU, fast link) trains fewer layers.}}
    \label{fig:splitting-example}
    \Description[]{}
\end{figure}

Federated Split Learning (FSL) \replaced{has been introduced as an alternative paradigm that}{advocates for a different approach: clients train the initial layers of a neural network, offloading the remaining layers to the server. This} directly alleviates client-side \replaced{resource}{memory and compute} consumption~\cite{Thapa2022-nt,Radovic2025-jc} while preserving, and in some cases\replaced{ even}{} enhancing~\cite{Thapa2022-nt},\replaced{model}{} accuracy. \replaced{By offloading a portion of the model to the server, FSL reduces the computational and memory requirements on clients, thereby enabling}{While FSL enables} a broader range of devices to participate in the training of larger\replaced{, potentially more accurate}{} models~\cite{Radovic2024-do}\replaced{. In FSL, clients train only the initial layers of a neural network, while the server trains the remaining layers, with training proceeding in a pipeline-parallel fashion~\cite{narayanan2019pipedream}.}{}\replaced{ However,}{,} training the same number of layers on all clients does not necessarily yield an efficient or fair distribution of computation, because such \replaced{static}{uniform} partitions ignore the underlying device heterogeneity.

To address this limitation, recent work~\cite{Samikwa2022-fr,Samikwa2024-ia,Mahmoud-Sajjadi-Mohammadabadi2025-pg} has explored adaptive FSL in which more powerful clients train deeper portions of the model, while weaker clients train fewer layers, as illustrated in \Cref{fig:splitting-example}. While more efficient than \replaced{static}{uniform}\replaced{ partitioning}{} schemes~\cite{Trepac2025-xv}, these approaches currently exhibit severe limitations. First, \replaced{their formulations largely}{they} adopt a client-centric perspective, implicitly assuming that the server always has sufficient\replaced{ computational and networking}{ compute and network} resources. In practice, however, overloading the server can shift the\replaced{ system}{} bottleneck from clients to the server itself, effectively making the server a straggler. Second, once a split \replaced{configuration }{}is assigned, clients are required to complete training with the allocated \replaced{number of }{}layers, despite the fact that computing resources and network connectivity of mobile devices are inherently dynamic. For instance, an initially well-balanced allocation may become suboptimal due to client mobility. Third, many solutions rely on explicit client-side profiling to estimate training time and network conditions, requiring preliminary procedures that consume resources without contributing to model convergence.

\replaced{In this work, w}{W}e argue that \emph{efficient and fair FSL requires a holistic, system-level perspective that accounts for both client- and server-side constraints in an online manner}: rather than committing to a split decided from a stale, upfront snapshot, the system needs to react \replaced{to}{as} conditions \replaced{as they }{}change during training. This approach reframes the split-point problem from a per-client, one-shot decision into a population-level, continuous control problem.
To this end, we propose \dynsl, a framework that dynamically distributes computation between clients and the server by minimizing the end-to-end training time. \replaced{Rather than relying on upfront measurements of client compute and network capabilities, \dynsl continuously monitors client-side conditions, server processing time, and effective network bandwidth during training by seamlessly hooking into the training mechanisms. Based on these measurements, it periodically solves a lightweight optimization problem to determine the split configuration that minimizes training time and, when the current configuration is suboptimal, \emph{immediately} reconfigures the system.}{}\replaced{In doing so, \dynsl quickly adapts to various system dynamics, such as fluctuating client compute capacity and network bandwidth, while maintaining efficiency and fairness across heterogeneous participants.}{\dynsl continuously tracks client compute and network capabilities and when necessary, it reconfigures the system to adapt to the changing system dynamics.} Rather than targeting a specific \replaced{algorithmic }{}configuration, we design \dynsl as an extensible framework that exposes configurable design choices\replaced{, including the number of classification heads to be trained on the server, whether client batches are concatenated, and the client sampling strategy.}{, such as whether client batches are concatenated and how many copies of the server-side model are trained.\footnote{We refer to each such copy (i.e., the entire server-side suffix of the model) as a classification head.}}
Overall, we make the following contributions:

\begin{enumerate}[nosep, leftmargin=*, label=(\arabic*)]
    \item We formulate a novel optimization problem that minimizes end-to-end FSL training time by jointly modeling client- and server-side costs -- the first split-allocation formulation to treat the server as a potential bottleneck -- and prove that it is solvable in polynomial time.
    \item We introduce within-epoch reconfigurations, immediately moving layers in response to runtime dynamics\replaced{ without waiting for an epoch to complete}{}. We further design a proportional aggregation scheme and prove it correctly accounts for each sample despite mid-epoch layer migration.
    \item We design a profiling-free online metric collection system that hooks into the training loop, eliminating wasted client compute and stale estimates of profiling-based methods.
    \item We evaluate \dynsl on a 24-device real-world heterogeneous testbed and in simulation, cutting training time by up to $30\%$ over the best baseline and $>\!50\%$ over a\replaced{n uninformed static}{ suboptimal uniform} split, with no loss -- and sometimes a gain -- in final accuracy.
\end{enumerate}
Ultimately, \dynsl fosters a more inclusive training process in which each participant -- server and clients alike -- contributes compute proportionally to its runtime capacity\replaced{, so that low-resource devices are neither excluded nor forced to straggle}{}. Our implementation is available at \replaced{\url{https://anonymous.4open.science/r/dynsl-3656/}}{\url{https://github.com/BorisRado/dynsl}}.

% The resulting efficiency gains improve every metric we measure: epoch time, time-to-accuracy, and in some cases final accuracy. 
%\rc{this version is much better than the last one.  would be nice if this paragraph can be further enriched a bit.}\iea{I agree. Where is the trade-off? Earlier we mention previous approaches have performance issues due to client heterogeneity etc., so where do we stand? Do we make gains? Do we lose a lot? Is our prediction performance loss acceptable for other benefits?}\borisinline{Better now? I'll include the results for accuracy today, but in short there is no degradation in accuracy; there's just a improvement in epoch time and time-to-accuracy. As per trade-off: of course this system is more complicated than training the same number of layers on all clients, but I don't think this should be mentioned here. Note I also added an impact statement below. Please remove the comments if the paragraph is now ok.}
% In summary, we make the following contributions:

% Ultimately, \dynsl fosters a more inclusive distributed learning process, wherein every participant (i.e., the server and every client) contributes data and compute proportionally to its availability and capacity. The consequent increase in efficiency benefits all metrics we analyze, including the epoch time, the time-to-accuracy, and in some cases, even the final accuracy. Our implementation is available at \url{https://anonymous.4open.science/r/dynsl-8F1D/}.
\section{Background and Preliminaries}
\label{sec:background}

\subsection{Split Learning}
\label{subsec:computational-graph-offloading}

To reduce the compute and memory requirements during training on resource-limited edge devices, Split Learning (SL)~\cite{Gupta2018-ff,Vepakomma2018-mb,Radovic2025-jc} proposes \emph{offloading} part of the computational graph to a server~\cite{Kang2017-qv}. Consequently, clients train only the initial layers of the model, while the server trains the remaining layers. During training, every client first performs a forward pass for a batch of data through its portion of the model and produces intermediate embeddings. These embeddings and the corresponding labels are sent to the server, which completes the forward pass, computes the loss and then performs backpropagation through its part of the model. This enables the server to update its model parameters and to obtain the gradients with respect to the client embeddings. Such gradients are finally sent to the client to complete the backward pass~\cite{Gupta2018-ff,Vepakomma2018-mb}. Since clients and server only exchange embeddings and gradients, SL provides some degree of privacy for client data%
%as client data cannot be fully reconstructed
~\cite{Abuadbba2020-fx,Deng2023-qp,XuUnknown-ld}. However, such a basic SL process only involves one client training at a time, motivating the need for federated split learning~\cite{Thapa2022-nt}.

\subsection{Federated Split Learning}
\label{subsec:sl_protocol}

Federated Split Learning (FSL) generalizes SL by enabling multiple clients to train concurrently. Similar to FL, FSL algorithms improve a model through a sequence of \emph{epochs}, which continue until a predefined stopping condition is met. Each epoch consists of the following stages~\cite{Radovic2025-jc}:
\begin{enumerate}[nosep, leftmargin=*, label=(\arabic*)]
    \item \emph{Client initialization:} A subset of clients is sampled from the population of available clients to participate in the current epoch. For each selected client, the server determines the \emph{split point}, i.e., the number of layers the client will train. This may be fixed in advance~\cite{Thapa2022-nt} or determined based on available information, such as previously estimated client training speed. Clients may have the same \replaced{split point, or a different split point may be assigned to each client}{or different split points}.
    \item \emph{Server initialization:} The server instantiates one or more classification heads, i.e., segments comprising the upper layers of the model that will be trained on the server in the epoch in conjunction with client-side ones.
    \item \emph{Model training:} Each client in parallel trains its assigned layers while offloading part of the computation to the server following the procedure presented in \secref{subsec:computational-graph-offloading}.
    \item \emph{Model aggregation:} After all clients complete training and upload their updated model parameters, the server aggregates all the collected trained weights in a layer-wise fashion, thereby obtaining the model for the next epoch. That is, if a layer has been trained locally by some clients and on the server by others (e.g., $\theta^3$ in \Cref{fig:splitting-example}), all client layers and the corresponding server layer are aggregated, e.g., using weighted averaging~\cite{McMahan2017-qj}.
\end{enumerate}

%An epoch completes only after the slowest participating client finishes uploading its updated parameters}.  

\section{Motivational Observations}
\label{subsec:motivation}

A key design decision in FSL -- and the aspect we focus on -- is determining \emph{how many layers each client should train}. This decision directly affects the computational load on both clients and the server, since the number and characteristics of the layers, such as their type and size, influence the training time as well as the volume of data exchanged over the network.  While prior work recognizes the importance of this choice~\cite{Samikwa2022-fr}, existing approaches share a common structure: they treat the split as a \emph{one-shot, client-centric decision derived from an upfront snapshot} of system conditions. Concretely, they assume (i) that the split can be chosen \emph{per client}; (ii) that a \emph{single} allocation suffices for an entire epoch; and (iii) that an upfront \emph{snapshot} of system conditions captures what the decision needs. We examine each assumption in turn and show that all three break down under realistic conditions.

% While prior work recognizes the importance of this choice~\cite{Samikwa2022-fr}, existing approaches share a common structure: they treat the split as a \emph{one-shot, client-centric decision derived from an upfront snapshot} of system conditions. Concretely, they assume \emph{(i)} that the split can be chosen per client, \emph{(ii)} that a single allocation can be appropriate for an entire epoch, and \emph{(iii)} that an upfront \emph{snapshot} of system conditions captures the information the decision needs. We next examine each of these assumption.

\borisparagraph{Is the split really a per-client decision?} Deciding the split \emph{per client} implicitly presupposes that the server can absorb whatever load clients offload onto it. This assumption holds when few clients participate, but as the population grows, the server itself can become the dominant bottleneck that caps training throughput~\cite{Trepac2025-xv}.

% Existing methods assume that only clients can become bottlenecks and that the server has sufficient computational capacity to never slow down training. In contrast, we argue that if the number of participating clients grows large, server-side resources can become the dominant bottleneck that limits training throughput 

To support this claim, we consider training a ResNet18 model on a set of Orange Pi (OP) devices and a server with either an A100 or P100 GPU using the SplitFed-v1 algorithm~\cite{Thapa2022-nt}. For this experiment, we emulate the actual epoch time by having multiple virtual clients generate embeddings at a pace characteristic of OPs, while the server performs the same tasks as if the clients were distributed physical devices. We evaluate four split points, specifically by training $4$, $5$, $6$, and $7$ layers on the client, corresponding to splitting the model before the first, second, third, and fourth residual blocks, respectively. Since OPs lack hardware accelerators, a natural intuition, exploited by several prior approaches~\cite{Shin2023-uy,Samikwa2022-fr,Wu2022-mz,Mahmoud-Sajjadi-Mohammadabadi2025-pg,Li2024-xm,Wazzeh2025-tr}, suggests training as few layers as possible on such clients. However, this intuition neglects the impact of offloading a large fraction of the model to the server.
% \footnote{In SplitFed-v1, each client maintains a separate classification head; as a result, the choice of split point does not affect convergence~\cite{Radovic2025-jc}.}

\begin{figure}[tbp]
    \centering
    \includegraphics[width=\linewidth]{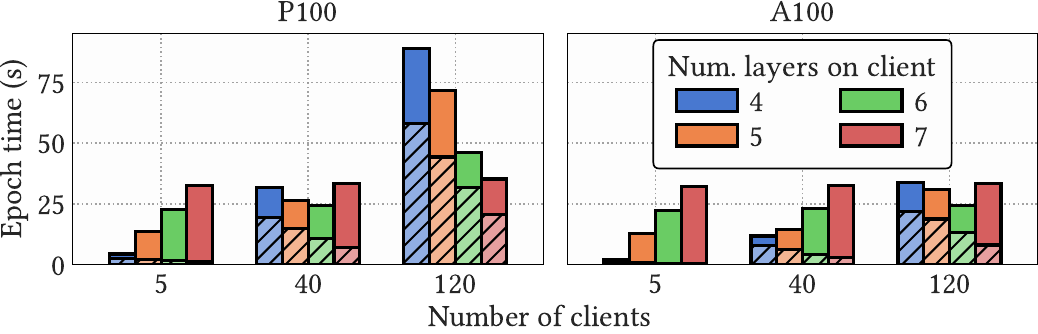}
    \caption{Epoch time (solid bars) and time during which the server GPU is actively training (bars with oblique hatches) when training ResNet18 with SplitFed-v1.}
    \label{fig:motivation_epoch_time}
    \Description{}
\end{figure}
In \Cref{fig:motivation_epoch_time}, we observe that, with $5$ clients, training as few layers as possible on clients yields the shortest epoch time. With $40$ or $120$ clients, however, the optimal split point depends on the server GPU. For instance, with $40$ clients, the best times are attained at $6$ client layers on the P100 and $4$ on the A100. Thus the \emph{same} client, on the \emph{same} hardware, should train a \emph{different} number of layers depending on (i) how many peers it contends with for server resources and (ii) the type of the server GPU. Any per-client rule that ignores this coupling cannot track the optimal split as the client population or the server changes, and will therefore misallocate work.
% \begin{observation}
% \label{obs:1}
%     In FSL, the optimal number of layers assigned to clients cannot be determined solely on a per-client basis, but must account for the population of concurrently training clients and the available server resources. 
%     % \iea{perhaps moving this observation text above just before the paragraph with 'existing methods assume' will alleviate that connection problem. Or leave this here and add some connecting text to the beginning of that paragraph.}
% \end{observation}
\Observation{obs:1}{In FSL, the optimal number of layers assigned to clients cannot be determined solely on a per-client basis, but must account for the population of concurrently training clients and the available server resources.}

\borisparagraph{Can a single split last a whole epoch?} Establishing the right split for a given population is not enough, however, because edge devices' conditions are inherently dynamic~\cite{Mahmoud-Sajjadi-Mohammadabadi2025-pg,Samikwa2022-fr,Bonawitz2019-zg}: client compute may fluctuate due to hardware-level mechanisms such as dynamic frequency scaling or user-driven events (e.g., the launch of background applications)~\cite{benoit2020impact}, and network conditions may vary because of client mobility, contention with co-located devices, or changes in wireless channel quality~\cite{MEI2022108736,mohan2019two}.

\begin{figure}[t]
    \centering
    \includegraphics[width=\linewidth]{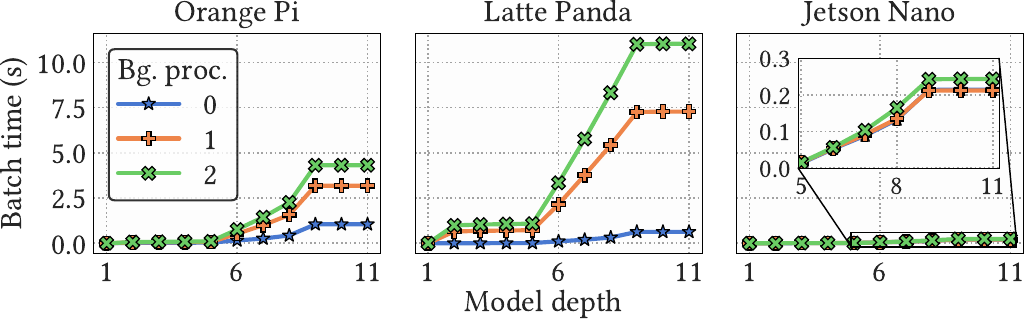}
    \caption{\replaced{Per-batch training time under varying number of background processes.}{Training time under varying background loads.}}
    \label{fig:time_wrt_bg_procs}
    \Description{}
\end{figure}

To quantify the variations in compute speed, \Cref{fig:time_wrt_bg_procs} reports the per-batch training time of three edge devices when training ResNet18 with zero, one, or two compute-intensive background processes. While the Jetson Nano mitigates additional load thanks to its GPU, the Orange Pi and Latte Panda exhibit substantial slowdowns even with a single background process: when training the full model, the per-batch \replaced{training }{}time increases by up to $3\times$ on the Orange Pi and $10\times$ on the Latte Panda. A shift of this magnitude is enough to turn a previously well-balanced client into a straggler mid-epoch, and because an epoch completes only when its slowest client finishes, \emph{a single straggler delays everyone}.\footnote{\replaced{This is not a rare case but a near-certainty at scale. If the probability that any given client experiences a slowdown during an epoch is just $1\%$, the probability that \emph{at least one} slows down is $\mathord{\sim}63\%$ with $100$ clients and $\mathord{\sim}87\%$ with $200$ clients.}{Slowdowns are likely at scale: if each client independently slows down with probability $1\%$, \emph{at least one} does with probability $\mathord{\sim}63\%$ for $100$ clients and $\mathord{\sim}87\%$ for $200$.}} A method that commits to a fixed split for the duration of the epoch has no recourse\replaced{ when this happens}{}.

% \begin{observation}
% \label{obs:2}
%     Mid-epoch slowdowns \replaced{are near-certain at scale}{become increasingly likely as the population grows}; keeping the split fixed when they occur may turn a client into a straggler, degrading training efficiency.
% \end{observation}
\Observation{obs:2}{Mid-epoch slowdowns \replaced{are near-certain at scale}{become increasingly likely as the population grows}; keeping the split fixed when they occur may turn a client into a straggler, degrading training efficiency.}

\borisparagraph{Can upfront measurements keep pace with runtime dynamics?} A natural response to device dynamics involves \replaced{re-}{}measuring client capabilities more often, and indeed existing approaches rely on \emph{profiling} phases performed periodically~\cite{Samikwa2022-fr,Samikwa2024-ia} or upon first participation~\cite{Mahmoud-Sajjadi-Mohammadabadi2025-pg}: clients train on synthetic data to measure layer-wise execution times and run additional procedures to estimate network throughput. However, this does not \replaced{re}{}solve the problem. Profiling captures only a snapshot, which quickly becomes outdated in dynamic environments, and profiling more frequently still lags behind real-time fluctuations while consuming client computation, energy, and time. Crucially, none of this effort contributes to model convergence: it is spent purely to collect control signals. \replaced{Coarser alternatives -- e.g., hardware specifications such as clock speed or core count~\cite{Wazzeh2025-tr} -- fail to capture runtime variability altogether.}{}
% \begin{observation}
% \label{obs:3}
%     Profiling-based decisions rely on potentially stale client measurements and incur non-negligible overhead.
% \end{observation}
\Observation{obs:3}{Profiling-based decisions rely on potentially stale client measurements and incur non-negligible overhead.}

\section{System Model}
\label{sec:system_setting}

% \iea{There is only one subsection in this section, so I don't see a point of keeping the subsection. Can this not be merged to the DynSL Design in the next section before the Framework Overview?}
% \borisnote{Removed the subsection. I'd keep setting \& \dynsl separate, but if you think it'd be better to merge them, I'll do it}
% We now formalize the system model by defining the participants, learning objective, and model structure. We also specify the assumptions governing client-server interaction and adversarial behavior.

\borisparagraph{System Setting.}
We consider a set of clients $C$, where each client $c \in C$ possesses a local dataset $D_c=\{(x_{c,k}, y_{c,k})\}_{k=1}^{|D_c|}$, which cannot be shared due to privacy constraints or regulatory reasons. An upstream entity provisions a compute server (e.g., on the cloud or on the edge) to assist clients during training by executing a portion of the model computation and coordinating the process.

\borisparagraph{Learning Objective.}
The goal of all participants is to collaboratively train a shared model $\theta$ that minimizes the average loss:
\begin{align}
    \theta^* = \arg\min_{\theta \in \mathbb{R}^d} 
        \frac{1}{|C|} \sum_{c \in C} 
        \mathbb{E}_{(x,y)\sim D_c}
        \big[ 
            \mathcal{L}(y, f(x,\theta))
        \big],
\end{align}
where $f(x,\theta)$ denotes the model output for input $x$ under parameters $\theta$, and $\mathcal{L}(\cdot)$ is a task-dependent loss function.

\borisparagraph{Model structure.}
The model $\theta=(\theta^1, \theta^2, \dots, \theta^L)$ is any deep neural network composed of $L$ sequential layers. Each layer $l$ consumes exclusively the output of its predecessor $l-1$, allowing the model to be expressed as a composition
\begin{align}
    f(x,\theta) = f\big(x, \theta^1 \circ \theta^2 \circ \dots \circ \theta^L\big).
\end{align}
Note that such layers need not correspond to layers in a deep learning model. Rather, they are the smallest part of the computational graph that cannot be further partitioned due to architectural constraints. For example, in ResNet models, each residual block is a layer because skip connections prevent finer splitting, and in transformers a layer corresponds to an entire transformer block.

\borisparagraph{Split computation.}
We study how to partition the model across heterogeneous clients in order to minimize training time. To this end, we introduce a control variable, referred to as the \emph{split vector},
\begin{align}
    \delta = \big[\delta_c\big]_{c \in C_i},
\end{align}
where $C_i \subseteq C$ denotes the set of clients participating in the training epoch $i$. For each client $c \in C_i$, the value $\delta_c \in \{0,1,\dots,L\}$ -- chosen by the server -- specifies the number of layers trained locally by that client. Concretely, $\delta_c = l$ indicates that client $c$ trains layers $1$ through $l$, while layers $l+1$ through $L$ are trained on the server. For instance, in \Cref{fig:splitting-example} we have $\delta=[2,3,3]$.

\replaced{This formulation subsumes several paradigms as special cases. When $\delta_c = L\,\,\forall c\in C_i$, the entire model is trained locally, recovering FL. When $\delta_c = 0 \,\,\forall c\in C_i$, no local training is performed and raw data is sent to the server, corresponding to centralized learning.}{}

\section{DynSL Design}

\replaced{We now present \dynsl, a framework that builds on the observations from \secref{subsec:motivation}. Following}{Building on observations from \secref{subsec:motivation}, we now present \dynsl . After} a high-level overview (\secref{subsec:framework-overview}), we describe how \dynsl relaxes each of the assumptions underlying state-of-the-art methods. First, it formulates a novel optimization problem that jointly accounts for client and server resources (\secref{subsec:optimization_problem_formulation}). Second, it hooks directly into the training loop\replaced{, ensuring that resource estimates remain up to date without incurring additional client-side overhead}{ to keep fresh estimates of client metrics (\secref{subsec:metric_collection})}. \replaced{Finally}{Third}, it introduces within-epoch reconfigurations \replaced{, which}{to} migrate layers between clients and the server (\secref{subsec:dynsl_workflow})\replaced{ without waiting for the current epoch to complete}{}.

\subsection{Framework Overview}%Framework 
\label{subsec:framework-overview}

\begin{figure*}[t]
    \centering
    \begin{subfigure}[t]{0.3\textwidth}
        \centering
        \includegraphics[width=\linewidth,page=1]{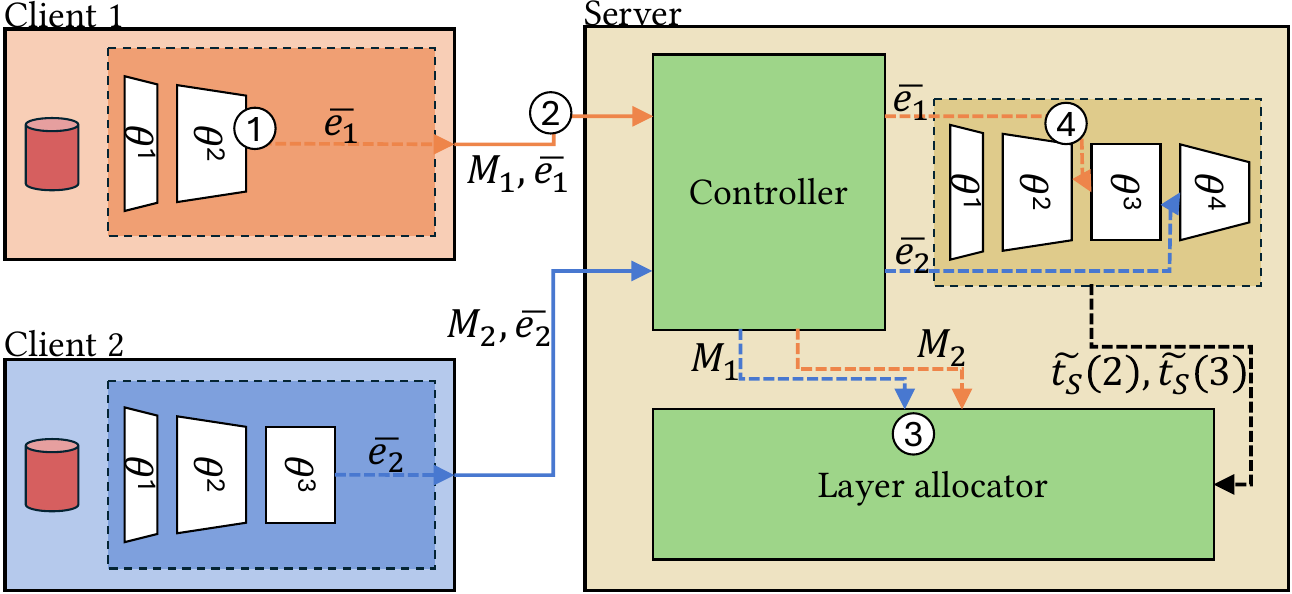}
        \caption{Regular training iteration.}
        \label{fig:x:a}
        \Description{}
    \end{subfigure}
    \hfill
    \begin{subfigure}[t]{0.3\textwidth}
        \centering
        \includegraphics[width=\linewidth,page=2]{images/dynsl_scheme_new.pdf}
        \caption{Moving layers from client to server.}
        \label{fig:x:b}
        \Description{}
    \end{subfigure}
    \hfill
    \begin{subfigure}[t]{0.3\textwidth}
        \centering
        \includegraphics[width=\linewidth,page=3]{images/dynsl_scheme_new.pdf}
        \caption{Moving layers from server to client.}
        \label{fig:x:c}
        \Description{}
    \end{subfigure}

    \caption{
    \replaced{
    \dynsl overview. (\ref{fig:x:a}): During training, \textcircled{1} clients perform a forward pass and \textcircled{2} send embeddings ($\bar{e}_1,\bar{e}_2$) and local metrics ($M_1, M_2$) to the server. The controller receives the data and \textcircled{3} forwards the metrics to the layer allocator and \textcircled{4} embeddings to the model for training. (\ref{fig:x:b}): When instructed to reduce the number of layers, \textcircled{1} the client removes the layers from the model, \textcircled{2} performs the forward pass through the reduced number of layers and \textcircled{3} sends the weights of the removed layers to the server along with the next batch of data. \textcircled{4} The server aggregates the weights of the client and server model, then it \textcircled{5} integrates metrics and \textcircled{6} trains the model. (\ref{fig:x:c}): When the layer allocator determines that the number of layers on a client is to be increased, \textcircled{1} it instructs the controller to \textcircled{2} extract the current state of the weights on the server, and \textcircled{3} send them along with gradients ($\bar{g}_2$). \textcircled{4} The client first performs the backward pass and then \textcircled{5} attaches the received layer to its model.
    }{
    \dynsl overview. (\ref{fig:x:a}): Clients \textcircled{1} run the forward pass and \textcircled{2} send embeddings ($\bar{e}_1,\bar{e}_2$) and metrics ($M_1, M_2$) to the server; the controller forwards \textcircled{3} the metrics to the layer allocator and \textcircled{4} embeddings to the model. (\ref{fig:x:b}): When instructed to reduce its layers, the client \textcircled{1} detaches them, \textcircled{2} runs the forward pass through the reduced model, and \textcircled{3} sends the removed weights with the next batch; the server \textcircled{4} aggregates the client and server weights, \textcircled{5} integrates the metrics, and \textcircled{6} trains the model. (\ref{fig:x:c}): To increase a client's layers, the allocator \textcircled{1} instructs the controller to \textcircled{2} extract the layer's current weights from the server and \textcircled{3} send them with the gradients ($\bar{g}_2$); the client \textcircled{4} completes the backward pass, then \textcircled{5} attaches the received layer.
    }
    }
    \label{fig:dynsl_overview}
    \Description[]{}
\end{figure*}

\dynsl adds a control loop atop standard FSL training. Two parties participate: (i) mobile/edge \emph{clients}, which train their assigned layers and stream runtime metrics to the server, and (ii) a compute-powerful \emph{server}, which hosts both the offloaded layers and \dynsl's control logic. This logic comprises a \emph{layer allocator} and a \emph{controller}. \Cref{fig:dynsl_overview} illustrates this loop; we detail each component next.

\subsubsection{Clients}
% Clients are edge devices that participate in training. Every sampled client receives an initial number of layers of the model and then enters the training loop, in which it performs a forward pass through its layers, sends the embeddings to the server, receives gradients, and completes the backward pass (\Cref{fig:x:a}).

Each sampled client runs the standard FSL training loop of \secref{subsec:computational-graph-offloading}. To support dynamic \emph{reconfigurations}, i.e., changing the split point during an epoch, training clients report their current runtime metrics (\secref{subsec:metric_collection}). When a reconfiguration happens, clients have to either increase or decrease the number of layers they train. In the former case, they receive the weights of the additional layers from the server and append them to their local model (\Cref{fig:x:c}), while in the latter case they remove the top layers from their model and send them to the server along with the next batch (\Cref{fig:x:b}).

\subsubsection{Server}

The server is a compute-powerful entity made up of one or more logically centralized nodes. It can be located in the cloud or at the network edge, where it assists nearby devices (e.g., for hierarchical FL~\cite{Fang2023-uy,lin2025hierarchical}). The server has three main components. 
% \rc{server can be a logical entity which actually consists of a set of hosts?}\borisinline{Clarified, hope it's ok now.}

\borisparagraph{Classification Heads.}
As in traditional FSL, \dynsl maintains one or more heads responsible for processing embeddings received from clients, completing the forward and backward passes, and returning gradients. To support reconfigurations, classification heads can provide the weights of selected layers to be transferred to clients (\Cref{fig:x:c}) and, when clients send weights to the server (\Cref{fig:x:b}), integrate such weights into their own.

\borisparagraph{Layer Allocator.}
The layer allocator determines the split vector $\delta$. At the beginning of each epoch, it assigns clients a pre-determined number of layers. During training, it continuously receives runtime metrics and, using such metrics, periodically solves the optimization problem described in \secref{subsec:optimization_problem_formulation} to identify the split vector that minimizes the epoch duration. If the resulting configuration differs from the current one, it instructs the controller to execute a reconfiguration.

\borisparagraph{Controller.}
The controller orchestrates communication between clients, classification heads, and the layer allocator. Upon receiving \replaced{embeddings from a client}{client data}, it forwards \replaced{them}{(a) embeddings} to the appropriate \replaced{classification }{}head along with an indication of how many layers the head needs to process -- note, for instance in \Cref{fig:x:a}, that the more layers the client has, the fewer layers the server trains for that client\replaced{. The controller also forwards}{ -- and (b)} client runtime metrics to the allocator\replaced{ and,}{.} Based on instructions from \replaced{it}{the allocator, it} handles reconfigurations. %Depending on the algorithm, the controller may also concatenate client batches.

\subsection{Optimization Problem Formulation}
\label{subsec:optimization_problem_formulation}

\dynsl minimizes the overall wall-clock training time by controlling the split vector $\delta$. Specifically, we minimize the duration $T_i$ of each epoch $i$ as, given a fixed number of epochs, minimizing each epoch's duration minimizes the total training time.\footnote{
\replaced{Minimizing wall-clock time is important to lower cost, as the server is typically a cloud GPU billed by provisioned time, and to speed up model delivery and testing.}{Minimizing time is key to lower cost and to speed up model delivery and testing.}}

As an epoch completes when the slowest client finishes its training stage, we seek to solve
\begin{align}
    \arg\min_{\delta} \; T_i(\delta)= \arg\min_{\delta} \; \max_{c \in C_i} T_{i,c}(\delta_c),
\end{align}
where $T_{i,c}$ denotes the \emph{actual} training time of client $c\in C$ in epoch $i$. We next derive expressions for \replaced{minimizing}{estimating    } the epoch time. Throughout the following discussion, we assume the system currently runs a configuration $\delta'$, from which \dynsl may transition to a new $\delta$.

\subsubsection{Client Perspective}

For a given number of layers $\delta_c$, client $c$ performs a forward and, upon receiving server gradients, a backward pass. The combined time at the client for the forward and backward passes depends on the split point and the client's compute capability $\psi_c$, and is denoted $t_c(\delta_c, \psi_c)$. The embeddings of size $|e_{\delta_c}|$ are uploaded to the server with bandwidth $\beta_c^{\uparrow}$, and the corresponding gradients of size $|g_{\delta_c}|$ are downloaded via bandwidth $\beta_c^{\downarrow}$.\footnote{We do not model latency because it can be amortized by overlapping communication and computation across multiple participating clients.} The server then completes the forward pass, computes the loss, and backpropagates it through its layers, which takes $t_S(\delta_c)$.

Beyond this per-batch cost, moving from $\delta'_c$ to $\delta_c$ requires migrating the affected layers over client $c$'s link, which takes time $m_c(\delta'_c, \delta_c)$. This cost is asymmetric because every layer a client trains must in any case be uploaded to the server at the end of the epoch. When the split is reduced ($\delta'_c > \delta_c$), the client sends to the server the layers it would have to upload regardless\replaced{. Therefore,}{, i.e.,} migrating them mid-epoch only brings this unavoidable transfer forward. When the split is increased ($\delta'_c < \delta_c$), instead, the additional layers must first be sent\replaced{ down}{} to the client and later returned, traversing the client link twice. Thus, if the size of the migrated layers is $P$, then:
\begin{align}
m_c(\delta'_c, \delta_c)=
\begin{cases}
0, & \delta'_c \ge \delta_c,\\
\frac{P}{\beta_c^{\downarrow}}+\frac{P}{ \beta_c^{\uparrow}}, & \delta'_c < \delta_c\replaced{,}{.}
\end{cases}
\end{align}
As per-batch stages occur sequentially while reconfiguration occurs only once, the total time for client $c$ to process its $B_c$ batches is
\begin{align}
    \hat T_{i,c}(\delta_c) = B_c \Big[
        t_c(\delta_c, \psi_c)
        + \frac{|e_{\delta_c}|}{\beta_c^{\uparrow}}
        + \frac{|g_{\delta_c}|}{\beta_c^{\downarrow}}
        + t_S(\delta_c)
    \Big] + m_c(\delta'_c, \delta_c).
\end{align}

\subsubsection{Server Perspective}

The client-centric view above, presented with some modifications in several prior works such as~\cite{Samikwa2022-fr,Mahmoud-Sajjadi-Mohammadabadi2025-pg,Wu2022-mz,Samikwa2024-ia,Deng2023-qp,Marinova2024-mf,Zhang2023-ef,Li2024-xm}, implicitly assumes that the server can process each batch immediately upon arrival. This assumption does not hold with many participating clients as discussed in \secref{subsec:motivation}. We thus characterize the epoch duration from the server's perspective.

Before an epoch completes, the server must process all batches from all participating clients. If it handles $b \geq 1$ batches in parallel (e.g., via concatenation or by training on multiple GPUs), the total server-side computation time is
\begin{align}\label{eq:server_comp}
    T_{S,\text{comp}}(\delta) = \sum_{c \in C_i} B_c \frac{t_S(\delta_c)}{b}.
\end{align}
In addition, the server exchanges embeddings and gradients for every batch. Assuming a symmetric link with bandwidth $\beta_S$, the total communication time is approximated as\footnote{We model the reconfiguration cost only on the client side because $\beta_S \gg \beta_c^{\uparrow}$\replaced{, thus the server clears these bytes faster than the client}{}.}
\begin{align}\label{eq:server_comm}
    T_{S,\text{comm}}(\delta) = \sum_{c \in C_i} B_c \frac{|e_{\delta_c}| + |g_{\delta_c}|}{\beta_S}.
\end{align}
Assuming the overlap between computation and communication, the server epoch duration is lower-bounded by
\begin{align}
\label{eq:server-time-max-compute-network}
    \hat T_S(\delta) \;=\; \max\!\Big(T_{S,\text{comp}}(\delta),\; T_{S,\text{comm}}(\delta)\Big).
\end{align}

\subsubsection{Estimated Epoch Time}

An epoch cannot complete until (i) all clients finish processing their local data, and (ii) the server finishes processing all client batches. We thus estimate epoch duration as
\begin{align}
\label{eq:epoch-time-final}
    T_i(\delta) \;\ge\; F(\delta) \;=\; \max\!\Big(\hat T_S(\delta),\; \max_{c \in C_i} \hat T_{i,c}(\delta_c)\Big).
\end{align}
% While $F(\delta)$ is a lower bound, it captures the dominant contributors to epoch duration and aligns with the empirical behavior observed in \secref{subsec:motivation}. In particular, early split points shift the bottleneck to the server, while split points deep in the model overload the clients.
While $F(\delta)$ is a lower bound, it is tight in its extremes: with a single client the server never queues, so epoch time equals the client term $F$ recovers exactly; at scale, the server sets the throughput ceiling, which dominates $F$. This also matches \Cref{obs:1}: early splits shift the bottleneck to the server, while deep splits overload clients.

\subsubsection{Problem Statement}

The layer allocator minimizes:
\begin{equation}
\label{eq:optimization_problem}
\boxed{
\begin{aligned}
& \underset{\delta}{\text{minimize}}
  & & F(\delta) \\
& \text{subject to}
  & & \delta_{\min}(c) \le \delta_c \le \delta_{\max}(c), \quad \forall c \in C_i\replaced{.}{,}
\end{aligned}
}
\end{equation}
where $\delta_{\min}(c)$ and $\delta_{\max}(c)$ denote client-specific bounds. The lower bound $\delta_{\min}(c)$ specifies the minimum number of layers a client trains locally to meet its privacy requirements -- deeper client-side models produce embeddings that are harder to invert~\cite{Abuadbba2020-fx,Deng2023-qp,XuUnknown-ld,pasquini2021unleashing}; a client may thus refuse to train very shallow parts. The upper bound $\delta_{\max}(c)$ caps the deepest split the client can sustain, capturing resource limitations such as available memory or providing an upper bound on energy usage. In short, \emph{$\delta_{\min}(c)$ encodes a privacy floor, while $\delta_{\max}(c)$ encodes a resource-consumption ceiling}.

This formulation lets \dynsl honor heterogeneous per-client preferences under a \emph{single} system-level objective. Rather than letting each client optimize its own objective -- which would yield $|C_i|$ selfish and one global throughput problem -- \dynsl adopts the $\epsilon$-constraint form: it minimizes a single global objective, the epoch time $F(\delta)$, and casts every other concern as a constraint on $\delta$.% Any metric monotone in the number of local layers maps directly onto these bounds---an energy or memory budget tightens $\delta_{\max}(c)$, a privacy requirement raises $\delta_{\min}(c)$---so a client tunes its preferences simply by adjusting its own bounds.

% This formulation lets \dynsl honor heterogeneous per-client preferences while retaining a \emph{single} system-level objective. A potential alternative exploited by related work is to let each client optimize its own scalarized trade-off between, e.g., time and energy. However, since the server needs to handle all participants (\Cref{obs:1}), this yields $|C_i|$ selfish problems plus one global throughput problem, with no principled weighting between individual and system-wide utility. \dynsl sidesteps this by adopting the $\epsilon$-constraint form of such multi-objective optimization: it minimizes a single global objective -- i.e., the epoch time $F(\delta)$ -- and expresses every other concern as a constraint on $\delta_c$. Whenever a metric is monotone in the number of local layers, it maps directly onto these bounds---an energy or memory budget tightens $\delta_{\max}(c)$, a privacy requirement raises $\delta_{\min}(c)$---so a client tunes its preferences simply by adjusting its own bounds.

\subsubsection{Solving the Allocation Problem}
% Finding solution to problem~\eqref{eq:optimization_problem}

While the number of possible split vectors is exponential, $O(L^{|C_i|})$, we show that under mild assumptions the problem is solvable in polynomial time.

\begin{assumption}\label{as:metrics}
All required metrics are available to the server.
\end{assumption}

\begin{assumption}\label{as:monotone}
Embedding sizes $|e_{\delta_c}|$ are non-increasing in $\delta_c$.
\end{assumption}
\noindent
Since gradients have the same dimensionality as the corresponding embeddings, $|g_{\delta_c}|$ is non-increasing as well; and since a deeper client split leaves fewer layers on the server, $t_S(\delta_c)$ is non-increasing too. Writing $\delta \preceq \tilde{\delta}$ whenever $\delta_c \le \tilde{\delta}_c$ for all $c \in C_i$, it follows that $\delta \preceq \tilde{\delta} \implies \hat T_S(\tilde{\delta}) \le \hat T_S(\delta)$. \Cref{as:monotone} holds for several architectures such as VGG, ResNet, and Transformers. When an architecture violates it, \dynsl restricts admissible splits to depths at which \replaced{$|e_l|$}{$|e_{\delta_c}|$} is a running minimum; this retains over $75\%$ of splits for EfficientNet and MobileNet we consider in \secref{sec:evaluation}.

\begin{claim}\label{claim:poly}
Under Assumptions~\ref{as:metrics} and~\ref{as:monotone}, problem~\eqref{eq:optimization_problem} can be solved \emph{exactly} in $O\big(|C_i|\,L\log(|C_i|\,L)\big)$ time.
\end{claim}

\begin{proof}[Proof sketch]
Fix the current configuration $\delta'$; then $\hat T_{i,c}(\cdot)$ is a fixed table of $O(|C_i|L)$ values, and by Assumption~2 the server time $\hat T_S(\cdot)$ is non-increasing in each coordinate. For a latency budget $\alpha$, let $\bar\delta(\alpha)$ assign every client its deepest feasible layer with $\hat T_{i,c}\le\alpha$. Among all admissible vectors whose client times are all $\le\alpha$, $\bar\delta(\alpha)$ is coordinatewise-largest and therefore, by monotonicity of $\hat T_S$, minimizes $\hat T_S$; its own bottleneck client time is $\le\alpha$. Consequently, any optimum $\delta^\star$ can be replaced by $\bar\delta(\alpha_0)$ with $\alpha_0=\max_c \hat T_{i,c}(\delta^\star_c)$ without increasing $F$: no client time exceeds $\alpha_0$ by construction, and $\hat T_S$ does not increase since $\delta^\star\preceq\bar\delta(\alpha_0)$. Hence an optimum lies in $\{\bar\delta(\alpha):\alpha\in\mathcal{A}\}$ with $\mathcal{A}=\{\hat T_{i,c}(l)\}$ and $|\mathcal{A}|\le|C_i|(L{+}1)$. We evaluate $F$ on these candidates by sweeping $\mathcal{A}$ in increasing order: consecutive budgets promote a single client \replaced{one layer deeper}{to a deeper split}, so $\bar\delta$ and the two server-time sums of Eqs.~\eqref{eq:server_comp}--\eqref{eq:server_comm} update in $O(1)$ per step. The sort dominates, giving $O(|C_i|L\log(|C_i|L))$.
\end{proof}

% \borisnote{An extended version of the proof is in the appendix; however, I'd keep the shorter version presented here}

The overall procedure, \Cref{algo:solving_optim_problem}, follows the proof: it sorts the candidate client times and sweeps them, evaluating $F$ at each vector and keeping the minimizer $\delta^\star$. As $F(\delta)$ is itself an approximate model of the epoch time, \dynsl\ does not act on arbitrarily small improvements: it triggers a reconfiguration only when $F(\delta')-F(\delta^\star)$ exceeds a hysteresis threshold $\xi$ -- an operator-configured margin (in seconds, e.g.,\ $5$-$10$\,s) set when the training job is launched.

\begin{algorithm}[t]
\caption{Solving problem~\eqref{eq:optimization_problem}}
\label{algo:solving_optim_problem}
\begin{algorithmic}[1]
\REQUIRE per-client bounds $[\delta_{\min}(c),\delta_{\max}(c)]$, current split $\delta'$
\STATE $\bar\delta_c\gets\arg\min_{l\in[\delta_{\min}(c),\delta_{\max}(c)]} \hat T_{i,c}(l)\ \ \forall c$
\STATE $\alpha_{\min}\gets\max_c \hat T_{i,c}(\bar\delta_c)$
\STATE $T_{S}^{\text{comp}}\gets T_{S,\text{comp}}(\bar\delta)$;\ \ $T_{S}^{\text{comm}}\gets T_{S,\text{comm}}(\bar\delta)$
\STATE $\delta^\star\gets\bar\delta$;\ \ $F^\star\gets\max \bigl(\max(T_{S}^{\text{comp}},T_{S}^{\text{comm}}),\,\alpha_{\min}\bigr)$
\STATE $\mathcal{P}\gets\textsc{sort}\bigl(\{(c,l):\delta_{\min}(c)\le l\le\delta_{\max}(c),\ c\in C_i\}\ \text{by}\ \hat T_{i,c}(l)\bigr)$
\FOR{$(c,l)\in\mathcal{P}$}
    \IF{$l > \bar\delta_c$}
       \STATE decrease $c$'s terms in $T_S^{\text{comp}},T_S^{\text{comm}}$;\quad $\bar\delta_c\gets l$
    \ENDIF
    \IF{$\hat T_{i,c}(l) \ge \alpha_{\min}$}
       \STATE $\hat F\gets\max\!\big(\max(T_S^{\text{comp}},T_S^{\text{comm}}),\,\hat T_{i,c}(l)\big)$
       \IF{$\hat F<F^\star$}
          \STATE $F^\star\gets \hat F$;\ $\delta^\star\gets\bar\delta$
       \ENDIF
    \ENDIF
\ENDFOR
\RETURN $\delta^\star$
\end{algorithmic}
\end{algorithm}

\subsection{Online Metric Collection}
\label{subsec:metric_collection}

Per \Cref{as:metrics}, to compute $F(\delta)$, the server must have estimates of (i) client-side computation time, (ii) client-server network bandwidth, (iii) server-side computation time, and (iv) the number of pending batches to be processed by each client. We next describe how \dynsl continuously tracks these metrics.

\subsubsection{Client Training Time}
\label{subsubsec:client-training-times}
Each client measures the wall-clock training time $\tilde{\tau}_c(l)$ of every local layer $l \leq \delta_c$ and piggybacks these values onto the embeddings sent to the server -- note that $t_c(\delta_c)=\sum_{l\leq \delta_c} \tilde \tau_c(l)$. As embeddings are orders of magnitude larger than the additional timing metrics, the overhead is negligible.

Upon receiving $\tilde{\tau}_c(l)$, the server updates the per-layer estimates using an exponentially weighted moving average (EWMA):
\[
\tau_c(l)
\leftarrow
\gamma \tau_c(l)
+
(1-\gamma)\tilde{\tau}_c(l),
\quad \forall l \leq \delta_c,
\]
where $\gamma \in [0,1)$. However, solving~\eqref{eq:optimization_problem} requires estimates for all split points $\delta_{\min}(c) < l \leq \delta_{\max}(c)$, including layers not currently trained by the client ($\delta_c < l \leq \delta_{\max}(c)$). To extrapolate beyond the observed depth, the server associates each layer $l$ with a complexity proxy $\kappa_l$ and predicts missing values via linear (affine) fit based on the observed layers, i.e., the server fits a regression model using observations $\kappa_l,\,\tau_c(l) \,\,\forall l \leq \delta_c$ and uses it to predict $\tau_c(l)\,\,\forall l >\delta_c$.

To validate this methodology, we quantify how accurately the per-client model predicts the training time of layers it has \emph{not} yet observed. Concretely, we adopt a leave-the-tail-out protocol that mirrors online use: for each device, we measure offline (for this validation only) the per-layer execution time at every depth, fit an affine model on the shallowest observed layers, and predict the execution time of deeper candidate splits.  In \Cref{tab:extrap}, we report the mean absolute percentage error (MAPE) of the resulting per-split estimate relative to the measured split cost for ResNet18; other models show consistent trends. Both FLOPs and every proxy based on measured execution time on a reference device (CPU or GPU) keep the error consistently below $22\%$ across all devices, confirming that a few observed layers suffice to extrapolate the cost of the rest\replaced{. In contrast, parameter counts extrapolate poorly. Note that the allocator does not require exact estimates, as it triggers reconfiguration only when the predicted improvement exceeds the hysteresis threshold~$\xi$. The residual error of the well-performing proxies is thus within the tolerance of the control loop, as confirmed by our end-to-end evaluation in \secref{sec:evaluation}.}{, whereas parameter counts extrapolate poorly. Such errors are tolerable for the allocator: extrapolated costs only cover layers a client has not yet trained and are replaced by measurements as soon as it does, while the threshold~$\xi$ prevents residual errors from triggering spurious reconfigurations.}
% , exceeding the other proxies by well over an order of magnitude, since parameter count is decoupled from actual compute. Their mean is further inflated by a long tail of extreme extrapolations (the median is around $100\%$), a skew the other proxies do not exhibit

\begin{table}[t]
\centering
\footnotesize
\caption{Relative error (\%) of the extrapolated per-split client compute time under leave-the-tail-out fitting. }
\label{tab:extrap}
\resizebox{\columnwidth}{!}{%
\begin{tabular}{l|rrrrrr|r}
% \begin{tabular*}{\columnwidth}{@{\extracolsep{\fill}} l|rrrrrr|r @{}}
\toprule
Proxy $\kappa_l$ & JAO & JN & JON & JXN & LP & OP & Mean \\
\midrule
CPU & 11.7 & 8.3 & 6.9 & 14.8 & 11.1 & 8.2 & \cellcolor[HTML]{ADD490}10.2 \\
A6000 & 6.3 & 8.3 & 7.4 & 21.5 & 7.0 & 10.2 & \cellcolor[HTML]{ADD490}10.1 \\
V100 & 17.4 & 15.0 & 16.0 & 18.1 & 14.3 & 14.7 & \cellcolor[HTML]{ADD490}15.9 \\
P100 & 9.2 & 7.8 & 8.6 & 19.3 & 7.0 & 7.9 & \cellcolor[HTML]{ADD490}10.0 \\
\midrule
FLOPs & 5.3 & 8.3 & 8.1 & 21.8 & 6.6 & 9.9 & \cellcolor[HTML]{ADD490}10.0 \\
Num. params. & 428.6 & 420.4 & 377.5 & 259.3 & 459.8 & 422.8 & \cellcolor[HTML]{F2D1BB}394.7 \\
\bottomrule
\end{tabular}}
\end{table}

We use the FLOPs proxy for the remainder of the paper: its precision is on par with the best execution-time proxies, yet unlike them it is computed directly from the model architecture and requires no reference-device measurement, making it the most practical choice. We stress, however, that \dynsl is modular here: any of the well-performing proxies can be substituted without altering the allocator, and their comparable precision indicates that the method does not depend on selecting a specific one.

\subsubsection{Client Network Bandwidth}

Rather than running dedicated profiling procedures, \dynsl observes that the continuous client-server exchange of embeddings and gradients already provides the data necessary for throughput estimation. We thus reuse the estimation principle of network profiling tools such as \texttt{iperf3} -- throughput as message size divided by the interval from the first payload byte to reception completion -- but apply it online during FSL training, at no additional networking cost. The client reports its downlink rate alongside the embeddings, while the server independently measures the client's uplink; both are smoothed with an EWMA across training steps to suppress transient fluctuations.

% Exchanged embeddings and gradients are sufficiently large to enable stable bandwidth estimation at batch granularity. For each message, the receiver measures the interval between the arrival of the first payload byte and the completion of reception, and estimates the ingress bandwidth as the ratio between message size (in bits) and this interval. The client reports its per-batch downlink bandwidth together with the embeddings, while the server independently measures its ingress rate (i.e., the client’s uplink bandwidth). To reduce transient fluctuations, measurements are smoothed using EWMA as training times.

\begin{figure}[tb]
    \centering
    \includegraphics[width=\linewidth]{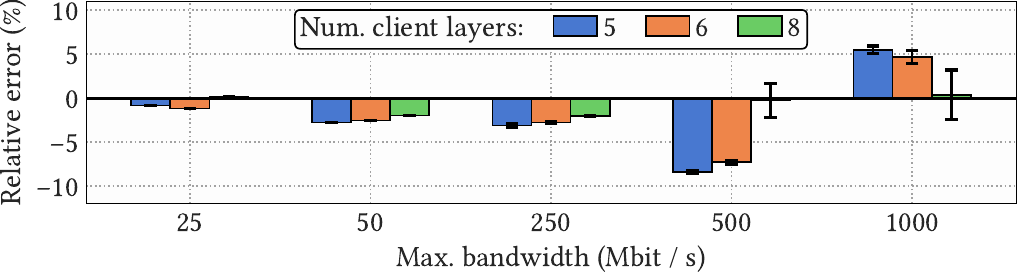}
    \caption{Relative error of bandwidth estimated by \dynsl relative to {\footnotesize\texttt{iperf3}} under various network regimes.}
    \label{fig:throughput_measurement}
    \Description[]{}
\end{figure}

We validate our protocol by estimating bandwidth while training ResNet18 on a single device under varying connection speeds. \Cref{fig:throughput_measurement} compares \dynsl's estimates against the \texttt{iperf3} utility: they track closely at low throughput and degrade slightly beyond ${\sim}500$~Mbps, yet the relative error never exceeds $10\%$. Crucially, unlike \texttt{iperf3}, \dynsl obtains these estimates without any dedicated profiling or added communication, and immediately reflects changing conditions as training proceeds.
% \borisnote{here I could emphasize more that we can not just use iperf because it would produce stale measurements.}

\subsubsection{Server-Side Metrics}

The server tracks its computation time and the number of pending batches for every client by decreasing a client-counter every time a batch is received. \replaced{}{Note that unlike client-side training times (\secref{subsubsec:client-training-times}), server-side times require no extrapolation: since each client $c$ starts at $\delta_\text{min}(c)$, the server initially trains the longest suffix the allocator may ever assign to it. Moreover, $t_S$ depends only on the split, not on the client, so these measurements are shared across clients.}

\vspace{0.5\baselineskip}

Altogether, by using such lightweight metric collection mechanisms, \dynsl maintains accurate, online estimates of all components required to evaluate $F(\delta)$. This design avoids explicit profiling, introduces negligible overhead, and enables rapid adaptation of the split configuration in response to dynamic system conditions.

\subsection{\dynsl Workflow}
\label{subsec:dynsl_workflow}

We finally describe the end-to-end workflow of \dynsl.

\borisparagraph{Client Registration.}
Clients may join at any time. Upon registration, a client reports (i) the size of its dataset, from which the server derives $B_c$, and (ii) the bounds $\delta_{\min}(c)$ and $\delta_{\max}(c)$.

\borisparagraph{Initial Layer Allocation.}
Since \dynsl performs no upfront profiling, the layer allocator \replaced{initializes conservatively, assigning each client $\delta_{\min}(c)$ at the start of the epoch.}{assigns each client $\delta_{\min}(c)$ at the start of every epoch, even when prior measurements of the client exist, as these may be stale (\Cref{obs:3}).}

\borisparagraph{Model Training and Dynamic Reconfiguration.}
Training follows the standard FSL protocol: the server processes client batches in the order they are received, optionally concatenating batches that offload the same number of layers into one. Throughout, it monitors system metrics and periodically runs \Cref{algo:solving_optim_problem} on the current estimates; when the resulting $\delta$ improves the estimated epoch time over the current $\delta'$ by more than \replaced{a threshold}{$\xi$}\iea{how is this threshold defined?}\borisinline{clarified above}, the controller triggers a reconfiguration. During training, both clients and the server maintain, for each layer, the number of samples on which it has been trained ($n^l_S$ and $n^l_c$, $\forall l \in [1,L], \forall c \in C_i$), as these counters are needed to merge weights correctly upon reconfiguration and at the end of the epoch. Because each client streams its dataset once per epoch, a layer migrated to a client mid-epoch is trained only on the samples the client has not yet processed. Two cases arise:

\begin{itemize}[nosep, leftmargin=*]
    \item $\delta_c > \delta_c'$: the controller extracts the affected layers from the server model and sends their weights to the client with the next gradient, recording the fork point $R^l_c \gets n^l_S$. The client appends the layers, sets $n^l_c = 0$, and trains them on its remaining batches.
    \item $\delta_c < \delta_c'$: the controller notifies the client alongside the gradient. After completing the backward pass, the client detaches the specified layers and returns their weights, together with their sample counts $n^l_c$, to the server, which integrates them immediately.
\end{itemize}

\borisparagraph{Weight Aggregation.}
At the end of the epoch, layers are combined by weighted averaging. However, a naïve FedAvg-style weighting would miscount samples when layers migrate between the server and clients during the epoch, as the resulting layers at the end of the epoch are not trained on disjoint sets of data -- the client's copy and the server copy share a common \emph{ancestor}.

\begin{figure}[h]
    \centering
    \includegraphics[width=0.9\linewidth]{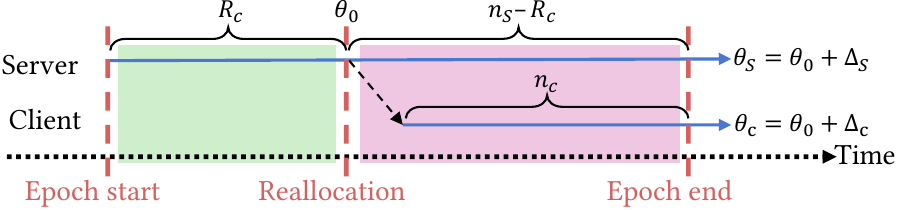}
        \caption{Example of migrating a layer within an epoch.}
    \label{fig:test}
\end{figure}
Consider the example in \Cref{fig:test}. A layer resides on the server, where it is trained on behalf of the whole population of clients. After $R_c$ samples, client $c$ \replaced{}{``}forks\replaced{}{''} it and trains it on its remaining $n_c$ samples, producing $\theta_c = \theta_0 + \Delta_c$. After the fork, the server copy keeps being trained by the remaining clients for $n_S - R_c$ additional samples, producing $\theta_S = \theta_0 + \Delta_S$. The two versions of the layer at the end of the epoch ($\theta_S$ and $\theta_c$) share the $R_c$-sample prefix baked into $\theta_0$ and their updates $\Delta_S$ and $\Delta_c$ are computed on disjoint sets of samples. Since both copies agree on $\theta_0$, the aggregation should, as in FedAvg, combine only the two divergent updates. However, naïve FedAvg-style weighting scores each copy by the \emph{total} number of samples it has seen -- i.e., $n_S$ for $\theta_S$ and $R_c + n_c$ for $\theta_c$ -- giving
\begin{align*}
    \theta_{\text{avg}}
    &\!=\! \frac{n_S\,\theta_S + (R_c+n_c)\,\theta_c}{n_S + R_c + n_c}\!=\! \theta_0
     \!+\! \frac{n_S}{n_S+R_c+n_c}\Delta_S
     \!+\! \frac{R_c+n_c}{n_S+R_c+n_c}\Delta_c.
\end{align*}
Both totals include the $R_c$ shared-prefix samples, which distorts the two updates' ratio. Specifically, the two disjoint updates are weighted $n_S : (R_c + n_c)$ rather than by the samples that actually produced them, $(n_S - R_c) : n_c$. The distortion is a bias toward whichever copy did \emph{less} post-fork work.

Two extreme cases show how far it can go. First, suppose the server copy is not updated after the fork ($n_S = R_c$, so $\Delta_S = \mathbf{0}$): the correct aggregate is simply $\theta_c$, yet naïve weighting still places weight $R_c / (2R_c + n_c)$ on the stale $\theta_0$, shrinking the client's genuine update $\Delta_c$ even though $\theta_S$ carries no information produced after divergence. Symmetrically, suppose the server keeps training on the whole population while the client contributes only a handful of post-fork samples: the $R_c$ prefix inflates the client's total to $R_c + n_c$, so its tiny tail is scored as if it rested on $R_c + n_c$ samples. With $R_c = 1000$, $n_c = 1$ and $n_S = 2000$, thus $n_S - R_c = 1000$, the naïve weighting still assigns the client weight $\tfrac{1001}{3001} \approx 0.33$, despite it having done $1000\times$ less post-fork work than the server.

To be consistent with FedAvg-style aggregation, layers should thus be weighted by how much data they have been trained on \emph{since \replaced{divergence}{forking}}. Aggregation must also account for clients forking at different points: the copies form a tree in which the server holds samples $1,\dots,n_S$, and each client branches off at its own $R_c$ and adds its $n_c$ samples.

Based on these considerations, we propose \Cref{algo:weight-aggregation}, which recombines the tree bottom-up, aggregating the client copies into the server copy one at a time in \emph{decreasing order of fork time} $R_c$ and weighting each merge by the samples accumulated since that fork. Processing the latest fork first ensures that, when a client is considered, the aggregate holds only that client's shared prefix plus samples disjoint from it, so the merge charges the ancestor for exactly the post-fork samples. Building on this, we characterize the model \Cref{algo:weight-aggregation} produces with \Cref{claim:agg}.\footnote{\replaced{The immediate integration performed mid-epoch follows the same principle and is omitted for brevity.}{Mid-epoch integration follows the same principles and is omitted for brevity.}}

\begin{algorithm}[t]
\caption{Weight aggregation for a single layer}
\label{algo:weight-aggregation}
\begin{algorithmic}[1]
\REQUIRE server weights $\theta_S$ trained on $n_S$ samples;
         client weights $\{\theta_c\}$, sample counts $\{n_c\}$, fork points $\{R_c\}$
\STATE $\theta_{\text{agg}} \gets \theta_S$, \quad $p \gets n_S$
\STATE sort clients in decreasing order of $R_c$
\FOR{each client $c$ in sorted order}
    \STATE $n_{\text{diff}} \gets p - R_c$
        \COMMENT{samples in the aggregate since the fork of $c$}
    \STATE $\theta_{\text{agg}} \gets
        \frac{n_{\text{diff}}}{n_{\text{diff}}+n_c}\,\theta_{\text{agg}}
        + \frac{n_c}{n_{\text{diff}}+n_c}\,\theta_c$
    \STATE $p \gets p + n_c$
\ENDFOR
\RETURN $\theta_{\text{agg}}$
\end{algorithmic}
\end{algorithm}

\begin{claim}\label{claim:agg}
\Cref{algo:weight-aggregation} maintains the invariant that, when client $c$ is merged, $n_{\mathrm{diff}} = p - R_c$ equals exactly the number of samples in $\theta_{\mathrm{agg}}$ trained after $R_c$, and these are disjoint from $c$'s $n_c$ samples. Hence each merge is a convex combination that weighs the two post-fork bodies in proportion to the samples that produced them, $n_{\mathrm{diff}}:n_c$.
\end{claim}

\begin{proof}
The claim follows once we establish, by induction on the merge order, the auxiliary invariant that $p$ always equals the number of distinct samples on which $\theta_{\mathrm{agg}}$ has been trained.

\emph{Base case.} Before any client is folded in, $\theta_{\mathrm{agg}} = \theta_S$ has been trained only on the server's $n_S$ samples and $p = n_S$, so $p$ counts the distinct samples on which $\theta_{\mathrm{agg}}$ has been trained.

\emph{Inductive step.} Suppose the invariant holds just before merging $c$. Because clients' contributions are aggregated in decreasing order of $R_c$, every client $c'$ considered before $c$ has $R_{c'}\! \ge\! R_c$, so its $n_{c'}$ samples, like the server samples processed after $R_c$, were trained after $R_c$. As client datasets are disjoint and each sample is processed exactly once per epoch, all of them are disjoint from $c$'s $n_c$ samples. Since no copy aggregated so far forked before $R_c$, the samples in $\theta_{\mathrm{agg}}$ are the shared prefix of exactly $R_c$ samples together with these post-$R_c$ samples; by the invariant the two groups total $p$, so there are $p\!-\!R_c = n_{\mathrm{diff}} \!\ge\! 0$ post-$R_c$ samples\rc{this sentence is unclear}\borisinline{please check again}. The merge $\frac{n_{\mathrm{diff}}}{n_{\mathrm{diff}} + n_c}\theta_{\mathrm{agg}} + \frac{n_c}{n_{\mathrm{diff}} + n_c}\theta_c$ is therefore convex and weighs the two disjoint bodies of $n_{\mathrm{diff}}$ and $n_c$ samples by their sizes. Finally, $p \gets p + n_c$ adds $c$'s $n_c$ fresh samples, so $p$ again counts the distinct samples in $\theta_{\mathrm{agg}}$, restoring the invariant.
\end{proof}
\noindent
The final model $\theta_{\mathrm{agg}}$ is thus the hierarchical FedAvg of the fork tree: each update is weighted by its post-fork sample count relative to its point of divergence. As a special case, when all clients fork at a common $R_c$, this reduces to FedAvg over post-fork samples. \replaced{}{Note, however, that \Cref{claim:agg} concerns sample accounting only: it does not make shared-head training equivalent to FedAvg, since server-side layers are updated more frequently than client-side ones and the dynamics further depend on the split and client batch arrival order.}

\section{Evaluation}
\label{sec:evaluation}

\borisparagraph{Comparison algorithms.}
We compare two variants of \dynsl against\replaced{ several}{} FSL baselines. For \dynsl, we consider the full algorithm as well as a simplified variant, denoted \dynsl-P, that performs \replaced{model }{}profiling and uses the collected measurements to solve (\ref{eq:optimization_problem}). Thus, in \dynsl-P, the optimal split vector is determined at the beginning of each epoch and remains fixed, while \dynsl initializes the clients with $\delta_\text{min}(c)$ and dynamically adapts \replaced{the split vector }{}$\delta$ during training. We group baselines by their effect on convergence. \emph{\ul{Convergence-preserving}} methods leave the computation unchanged, hence are FedAvg-equivalent under separate heads: SplitFed-v1~\cite{Thapa2022-nt} (varied client-side depth) and the resource-aware ARES~\cite{Samikwa2022-fr} -- we take ARES as a strong, representative \emph{server-blind} adaptive baseline. Other adaptive methods (see \secref{sec:related-work}) differ \replaced{only }{}in \emph{how} they select per-client splits, not in \emph{whether} they model server load; by Observation~\ref{obs:1} they therefore share the same structural failure mode, so \dynsl's advantage over ARES is representative of this family. \replaced{}{Note that under separate heads, ARES, \dynsl-P, and \dynsl form an ablation ladder: ARES$\rightarrow$\dynsl-P isolates server awareness, as both profile clients and keep $\delta$ fixed throughout the epoch, while \dynsl-P$\rightarrow$\dynsl isolates the combined effect of online adaptation and the profiling-free start.} \emph{\ul{Convergence-altering}} methods change what is computed: SFSL~\cite{Han_undated-pd} and DTFL~\cite{Mahmoud-Sajjadi-Mohammadabadi2025-pg} train clients on a locally generated loss via auxiliary networks rather than on server gradients. 

% For static split-point algorithms that assign the same number of layers to all clients, we evaluate SplitFed-v1 and SplitFed-v2~\cite{Thapa2022-nt} with varying numbers of client-side layers, as well as SFSL~\cite{Han_undated-pd}, an FSL method in which clients optimize a locally generated loss instead of relying on gradients from the server. Among resource-aware approaches, we consider ARES~\cite{Samikwa2022-fr} and DTFL~\cite{Mahmoud-Sajjadi-Mohammadabadi2025-pg}.

\borisparagraph{Datasets and models.}
We evaluate several model architectures: ResNet18~\cite{he2016deep}, EfficientNet~\cite{tan2019rethinking}, MobileNet~\cite{howard2017mobilenets}, DenseNet~\cite{huang2017densely}, and a small Vision Transformer model~\cite{dosovitskiy2020image}. We train these models on the CIFAR10, CIFAR100, and CINIC10 datasets. Depending on the experiment, we partition the underlying training set of the considered datasets using either an IID or Dirichlet distribution and assign the so-obtained subsets to clients.

% \borisparagraph{Datasets and models.}
% For each experiment, the training dataset is partitioned into $25$ client datasets under an IID setting. All clients are sampled in every epoch and the model performance is evaluated at the end of each training epoch using the corresponding test dataset. During training, and following common practice in the FSL literature~\cite{Gao2022-kv,Dachille2024-yx,Chopra2021-vv,Abuadbba2020-fx,Kim2023-wq,Li2021-zc,Poirot2019-ow,Titcombe2021-jm}, we use the Adam optimizer with a constant learning rate of $0.001$, and a batch size of $16$ on the clients.

\borisparagraph{Implementation and testing environment.}
We evaluate \dynsl on CoLExT~\cite{Bovzivc2024-ms}, a state-of-the-art real-world testbed consisting of the 24 heterogeneous devices listed in \Cref{tab:colext}; the server is equipped with an A6000 GPU and $48$ cores. For details refer to~\cite{Bovzivc2024-ms}. % Experiments in \secref{sec:varying-participation} and \secref{subsec:simulation-accuracy-different-datasets} are obtained through simulations.

\replaced{}{\borisparagraph{Algorithm Hyperparameters.} For \dynsl we set $\gamma=0.8$. The allocator runs every 8\,s, and reconfigurations are triggered when predicted gains exceed $\xi=8$\,s; the server processes one batch at a time ($b=1$). We use FLOPs as the complexity proxy $\kappa$. We set these values and keep them fixed across all experiments, so that no result benefits from per-experiment tuning. For ARES, we set parameter $\alpha$~\cite{Samikwa2022-fr} to $1.0$ so that ARES optimizes training time only.}

\begin{table}[b]
\centering
\footnotesize
\caption{Composition of the employed testbed.}
\label{tab:colext}

\begin{tabular}{@{}r|c|c|c|c|c@{}}
\toprule
                    & Acronym & Count & GPU    & Memory & CPU (Core@GHz)  \\ \midrule
Jetson AGX Orin     & JAO     &    2  & \cmark & 64GB & 12@2.2 \\
Jetson Orin Nano    & JON     &    3  & \cmark & 8GB  & 6@1.5  \\
Jetson Xavier NX    & JXN     &    2  & \cmark & 8GB  & 6@1.9  \\ 
Jetson Nano         & JN      &    6  & \cmark & 4GB  & 4@1.5  \\ 
Latte Panda         & LP      &    3  & \xmark & 8GB  & 4@2.9  \\ 
Orange Pi           & OP      &    8  & \xmark & 16GB & 8@2.3  \\ 
\bottomrule
\end{tabular}
\end{table}

\subsection{Training Time with Separate Heads}
\label{subsec:separate-clf-heads}

\replaced{Throughout this section we consider \emph{convergence preserving} algorithms (SplitFed-v1, ARES, \dynsl, \dynsl-P), wherein each client is paired with a separate classification head. Under this configuration, no modifications are made to the training algorithm -- i.e., training a layer on the server or the client does \emph{not} alter the actual computations being performed --, thus, the split point does not affect the convergence dynamics~\cite{Radovic2025-jc,han2024convergence}. Consequently, the per-epoch evolution of loss and accuracy is identical to that of FedAvg, aside from minor stochastic variations caused by random components (e.g., dropout) and finite numerical precision. As a result, \emph{reducing the epoch duration directly reduces the total training time without affecting the final model accuracy}.}{}

\replaced{}{We consider convergence-preserving algorithms (SplitFed-v1, ARES, DynSL, DynSL-P) and report the mean and standard deviation of the epoch time over eight epochs within a single run, excluding the profiling time of ARES and DynSL-P, which favors these baselines. Because each client is paired with a separate classification head, the split point only changes where a layer is trained, not what is computed, so convergence is equivalent to that of FedAvg~\cite{Radovic2025-jc,han2024convergence}. Reducing the epoch duration therefore directly reduces the total training time without affecting the final accuracy.}

% We evaluate both variants of \dynsl and compare them with algorithms, which do not alter convergence, namely ARES~\cite{Samikwa2022-fr} and SplitFed-v1~\cite{Thapa2022-nt} under different split-point configurations -- Early, Middle, and Late.

\subsubsection{Stable Client Conditions}

We report in \Cref{tab:epoch-time-heads} the epoch time in the ideal case, in which no client experiences compute or network slowdowns. Across all architectures and both data distributions, \replaced{the two \dynsl variants attain the lowest epoch times}{the lowest epoch time is attained by one of the \dynsl variants (\dynsl-P in 8 of 10 cells)}, reducing the duration by up to ${\sim}30\%$ relative to the best competitor (e.g., $80.1$s $\rightarrow$ $57.9$s for EfficientNet on IID data) and by more than $50\%$ compared to \replaced{fixed}{a uniform} split with suboptimal \replaced{choice}{depth} -- note that, as per \Cref{obs:1}, there is no way of knowing a priori which split point is the best. \replaced{\dynsl-P is the strongest single method, best in nearly every cell; in the few where it is not (e.g., MobileNet-IID), it is within standard deviation of the best.}{}

\begin{table}[t]
  \centering
  \caption{Epoch time in seconds with per-client heads under stable client conditions.}
  \label{tab:epoch-time-heads}
  \footnotesize
  \setlength{\tabcolsep}{4pt}
  \begin{tabular}{l | ccccc}
    \toprule
    Method & MobileNet & EfficientNet & ViT & ResNet18 & DenseNet \\
    \midrule
    \multicolumn{6}{l}{\emph{IID}}\\
    \dynsl{}-P & \meanstd[3]{48.3}{0.1} & \meanstd[1]{57.9}{3.0} & \meanstd[1]{92.0}{3.7} & \meanstd[1]{109.6}{1.1} & \meanstd[3]{126.3}{1.4} \\
    \dynsl{} & \meanstd[1]{46.7}{2.8} & \meanstd[2]{61.7}{1.8} & \meanstd[2]{100.2}{0.5} & \meanstd[2]{114.0}{2.3} & \meanstd[1]{125.5}{0.9} \\
    ARES & \meanstd{64.8}{4.3} & \meanstd{93.6}{1.1} & \meanstd{118.9}{0.3} & \meanstd[3]{138.0}{1.0} & \meanstd[2]{126.2}{0.6} \\
    SplitFed-Early & \meanstd{55.6}{0.9} & \meanstd{90.8}{1.5} & \meanstd[3]{117.3}{0.6} & \meanstd{146.7}{0.5} & \meanstd{218.4}{1.6} \\
    SplitFed-Mid. & \meanstd[2]{47.2}{0.8} & \meanstd[3]{80.1}{0.6} & \meanstd{137.5}{2.3} & \meanstd{145.5}{7.5} & \meanstd{146.2}{0.7} \\
    SplitFed-Late & \meanstd{55.9}{0.1} & \meanstd{123.9}{0.4} & \meanstd{211.4}{0.7} & \meanstd{209.7}{16.2} & \meanstd{146.1}{0.4} \\
    \midrule
    \multicolumn{6}{l}{\emph{Dirichlet ($\alpha=0.2$)}}\\
    \dynsl{}-P & \meanstd[1]{62.1}{0.1} & \meanstd[1]{91.3}{2.0} & \meanstd[1]{145.6}{1.6} & \meanstd[1]{154.9}{2.3} & \meanstd[1]{229.3}{0.2} \\
    \dynsl{} & \meanstd[2]{70.4}{1.0} & \meanstd[2]{97.5}{0.3} & \meanstd{172.5}{1.5} & \meanstd[2]{158.4}{9.2} & \meanstd[3]{245.6}{0.3} \\
    ARES & \meanstd[3]{88.1}{5.0} & \meanstd[3]{119.2}{0.9} & \meanstd[2]{170.6}{2.2} & \meanstd[3]{214.8}{4.0} & \meanstd[2]{234.7}{1.3} \\
    SplitFed-Early & \meanstd{88.2}{2.2} & \meanstd{122.1}{0.4} & \meanstd[3]{171.0}{1.0} & \meanstd{234.5}{3.2} & \meanstd{350.1}{5.3} \\
    SplitFed-Mid. & \meanstd{92.2}{0.3} & \meanstd{170.9}{0.5} & \meanstd{242.8}{0.9} & \meanstd{263.5}{1.6} & \meanstd{297.8}{1.1} \\
    SplitFed-Late & \meanstd{118.4}{0.1} & \meanstd{262.6}{0.5} & \meanstd{413.4}{1.0} & \meanstd{344.8}{12.3} & \meanstd{313.5}{1.9} \\
    \bottomrule
  \end{tabular}
\end{table}

\begin{figure}[t]
    \centering
    \includegraphics[width=\linewidth]{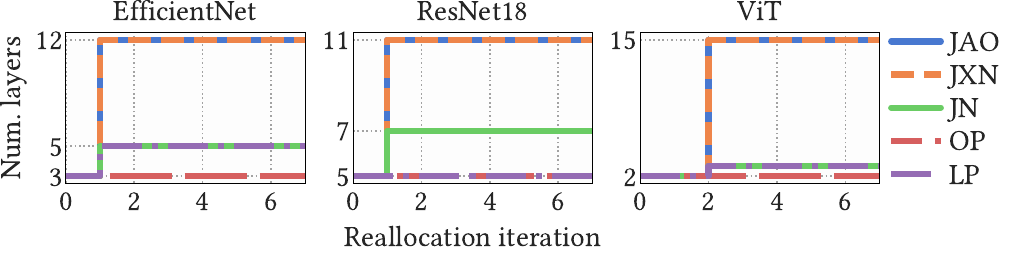}
    \caption{Allocated layers over an epoch with \dynsl.}
    \label{fig:layer-evolution-no-bg}
    \Description[]{}
\end{figure}

\dynsl recovers almost all of \dynsl-P's benefit \emph{without any profiling}. As \Cref{fig:layer-evolution-no-bg} shows, \dynsl starts every client at its minimum split -- shifting load to the server -- and progressively migrates layers outward as runtime measurements arrive. This cold start costs a short initial phase in a suboptimal configuration: under IID data the resulting gap to \dynsl-P stays within ${\sim}9\%$, widening under skew (up to ${\sim}19\%$ for ViT) where it occasionally trails ARES on the largest models. We view this as a favorable trade: the profiling-free \dynsl removes all client-side benchmarking overhead and stale estimates, yet still outperforms every \replaced{static}{uniform} baseline\replaced{ and matches the profiling variant once adapted}{ in 9 of 10 cells and trails by <1\% in the tenth}.

\begin{figure}[t]
    \centering
    \includegraphics[width=\linewidth]{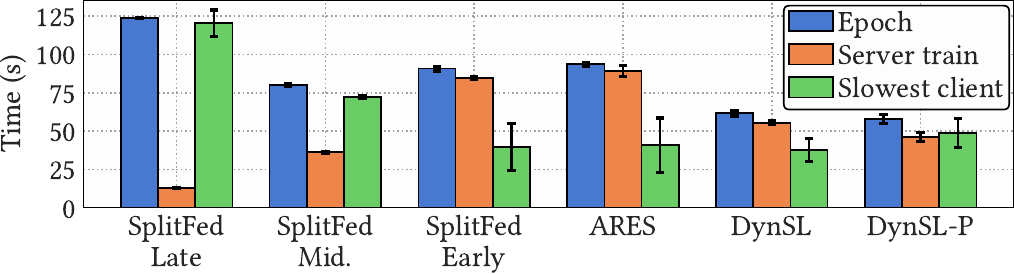}
    \caption{Breakdown of the epoch times when training EfficientNet with separate heads per client.}
    \label{fig:train-time-breakdown-no-bg}
    \Description[]{}
\end{figure}

To expose \emph{why}, \Cref{fig:train-time-breakdown-no-bg} decomposes the EfficientNet epoch into the server's active training time (``Server train'') and the slowest client's active time (``Slowest client''). Since the epoch length cannot be shorter than these two contributors (Eq. \ref{eq:epoch-time-final}), a method is fast only when it balances them. Both \dynsl variants sit at this sweet spot, whereas the baselines fall to one side or the other: SplitFed-``Late'' is client-bound (slowest client $\approx$ epoch), while SplitFed-``Early'' and ARES are server-bound (server train $\approx$ epoch). ARES's behavior is a direct symptom of \Cref{obs:1}: by omitting server contention from its formulation, it over-offloads, leaves clients training only shallow layers, and turns the server into the bottleneck -- in several cases yielding epoch times worse than a \replaced{fixed}{uniform} split. \dynsl instead accounts for both sides, keeping neither idle.

\subsubsection{Impact of Compute Slowdowns}

We stress a single client mid-epoch, slowing it down five seconds after the epoch begins in one of two ways: launching a compute-intensive background process on a CPU-only LP, or lowering the clock frequency of a GPU-equipped JN (as under thermal throttling or energy saving).\replaced{}{\footnote{We perturb a single client; the allocator re-solves the split for the whole population at each invocation, so the same mechanism applies to slowdowns of multiple clients.}} Both inflate the client's per-batch time (\Cref{obs:2})\replaced{; the only remedy is to adapt the affected client's split to the reduced capacity, which no method other than \dynsl can do.}{. A method can mitigate this either by reducing the affected client's split mid-epoch as \dynsl, or by assigning clients relatively little work upfront that the slowdown stays hidden, at the price of making the server the bottleneck.}

\Cref{tab:time-slowdowns} reports the resulting epoch times (IID data). \replaced{\dynsl is the single best method in every case. Comparing against the stable-condition baseline of \Cref{tab:epoch-time-heads}, it is also the only method whose epoch time is essentially unchanged by the slowdown (EfficientNet $61.7\!\rightarrow\!62.4/64.5$\,s; ViT $100.2\!\rightarrow\!98.7/96.9$\,s)}{\textsc{DynSL} is the fastest method in every case, cutting epoch time by 10--33\% relative to the next-best method, and stays within 5\% of its stable-condition time from Table~\ref{tab:epoch-time-heads} (EfficientNet 61.7$\to$62.4/64.5\,s; ViT 100.2$\to$98.7/96.9\,s)}. \dynsl-P, on the other hand, while being the fastest method under stable conditions, under a mid-epoch slowdown collapses to among the slowest (EfficientNet $57.9\!\rightarrow\!109.9$\,s; ViT $92.0\!\rightarrow\!159.9$\,s), unable to \replaced{revise}{cope with} the medium-to-deep split it assigned the now-slow client. Among the static baselines, configurations that keep clients shallow (SplitFed-Early,\replaced{ and likewise}{} ARES) barely move, since the slowdown is experienced by a client training only a shallow part of the model; they remain overall slow, however, as they push most of the computation onto the server. Configurations that load the clients (SplitFed-Mid./Late) inflate instead, worst for SplitFed-Late. \dynsl avoids both failure\replaced{ mode}{}s: rather than committing to a split upfront, it revises each client's depth from live measurements, so when a client slows down it immediately offloads that client's layers to the server -- reducing its depth potentially \replaced{up}{down} to the minimum feasible $\delta_{\min}$ -- removing it from the critical path.

\begin{table}
\centering
\footnotesize
\caption{Epoch time under a mid-epoch compute slowdown.}
\label{tab:time-slowdowns}
\begin{tabular*}{\columnwidth}{@{\extracolsep{\fill}} l|cc|cc}
\toprule
 & \multicolumn{2}{c|}{Clock throttling (JN)} & \multicolumn{2}{c}{Bg. process (LP)} \\ 
 & EfficientNet & ViT & EfficientNet & ViT \\
\midrule
DynSL-P & \meanstd{109.9}{0.3} & \meanstd{159.9}{0.3} & \meanstd[2]{71.7}{2.3} & \meanstd{119.6}{27.7} \\
DynSL & \meanstd[1]{62.4}{0.3} & \meanstd[1]{98.7}{5.7} & \meanstd[1]{64.5}{1.1} & \meanstd[1]{96.9}{2.1} \\
ARES & \meanstd[3]{93.2}{1.1} & \meanstd[3]{112.9}{0.4} & \meanstd{90.5}{0.4} & \meanstd[2]{112.1}{0.7} \\
SplitFed-Early & \meanstd[2]{92.9}{0.2} & \meanstd[2]{112.6}{0.5} & \meanstd[3]{90.4}{0.0} & \meanstd[3]{112.5}{0.3} \\
SplitFed-Mid. & \meanstd{108.7}{0.3} & \meanstd{174.3}{35.6} & \meanstd{94.5}{9.4} & \meanstd{197.0}{37.6} \\
SplitFed-Late & \meanstd{186.3}{0.2} & \meanstd{202.5}{1.6} & \meanstd{132.9}{10.7} & \meanstd{201.6}{2.7} \\
\bottomrule
\end{tabular*}
\end{table}

\subsubsection{Impact of Bandwidth Degradation.} We isolate the communication axis by measuring how reduced network throughput affects epoch time. We thus train the ResNet18 model on CINIC10 and, five seconds after the start of each epoch, throttle both the ingress and egress bandwidth of a single OP client using the Linux \verb|tc| traffic-control utility, emulating the throughput drop a client experiences under weak coverage or network contention.\footnote{
\replaced{CoLExT by default provides a shared 1~Gbps client--server link, representative of high-capacity edge links such as 5G under ideal conditions~\cite{rochman2025comprehensive,xu2020understanding}. Such conditions need not persist during training: a client on a fast, uncongested link can see its throughput fall to tens or hundreds of Mbps as it moves toward weaker coverage or as other devices share the link~\cite{narayanan2020first}, motivating the mid-epoch drop we emulate.}{CoLExT's default shared 1~Gbps client--server link matches 5G under ideal conditions~\cite{rochman2025comprehensive,xu2020understanding}; with weaker coverage or link sharing, throughput can fall to tens or hundreds of Mbps~\cite{narayanan2020first}, which our mid-epoch drop emulates.}
} We then measure the resulting epoch completion time as a function of the client-observed throughput for \replaced{}{ARES, }SplitFed, \dynsl, and \dynsl-P.

Table~\ref{tab:reduced-throughput-tests} reports epoch time with respect to the throttled client's throughput. Two facts stand out. First, the best static configuration changes with throughput. On an unthrottled link, aggressive offloading (\dynsl-P, SplitFed-Early) is fastest, but this inverts below 50~Mbps: these configurations assign the throttled OP a shallow split, so the client must exchange large embeddings. Consequently, such configurations become the \emph{slowest} as link quality degrades.

Second, \dynsl~is the only method that tracks this moving optimum. It is fastest at 10~Mbps ($3\times$ faster than \dynsl-P) and stays within $12\%$ of the per-regime best baseline elsewhere. Since \dynsl-P determines the split from an upfront snapshot and commits to it for the entire epoch, it cannot react to the degraded throughput, and its own epoch time inflates from 112 to 981s as throughput falls. Note that \dynsl~is not strictly best in every regime because, as discussed above, it starts in a suboptimal configuration and, when network quality changes, needs a few seconds to detect the change and reconfigure, during which it runs suboptimally.

\begin{table}
\centering
\footnotesize
\caption{Epoch time under mid-epoch network slowdown.}
\label{tab:reduced-throughput-tests}
\begin{tabular*}{\columnwidth}{@{\extracolsep{\fill}}lcccc@{}}
\toprule
& \multicolumn{4}{c}{Reduced throughput (Mbps)} \\
& 10 & 20 & 50 & Unthrottled \\
\midrule
DynSL-P  & \meanstd{981.3}{5.5} & \meanstd{525.4}{2.8} & \meanstd{259.4}{0.5} & \meanstd[1]{112.4}{2.4} \\
DynSL  & \meanstd[1]{322.4}{2.0} & \meanstd[2]{292.0}{2.6} & \meanstd[2]{211.2}{1.4} & \meanstd[2]{119.9}{2.7} \\
ARES  & \meanstd{984.2}{3.8} & \meanstd{530.2}{4.3} & \meanstd{260.3}{1.2} & \meanstd[3]{132.8}{2.3} \\
SplitFed-Early & \meanstd{979.1}{3.2} & \meanstd{525.9}{1.2} & \meanstd{255.7}{0.0} & \meanstd{148.5}{2.9} \\
SplitFed-Mid. & \meanstd[3]{566.2}{0.3} & \meanstd[3]{341.5}{0.7} & \meanstd[1]{205.6}{0.7} & \meanstd{151.6}{7.4} \\
SplitFed-Late & \meanstd[2]{381.9}{1.7} & \meanstd[1]{261.2}{0.6} & \meanstd[3]{213.6}{10.1} & \meanstd{217.4}{12.8} \\
\bottomrule
\end{tabular*}
\end{table}

%%%%%%%%%%%%%%%%%%%%%%%%%%%%%%%%%%%%%%%% 

\subsubsection{Robustness to Population Size and Composition}
\label{sec:varying-participation}
The testbed experiments above use full participation: one fixed client set trains every epoch. This does not exercise \dynsl's central premise (\Cref{obs:1}), i.e., that the optimal split depends on the concurrent \emph{population} and the server contention it induces. We therefore vary the population's size and composition via emulation, reusing the setup of \secref{subsec:motivation}: virtual clients produce embeddings at their device-characteristic rate while the server runs the actual offloaded workload, so the server contention central to \Cref{obs:1} is measured, not modeled. Client training time is thus modeled from per-device measurements; we emulate the network by idling each virtual client for a time set by its message size and simulated throughput. To confirm fidelity, we reproduce the full-participation configurations of the previous sections; emulated epoch times stay within 10\% of the CoLExT measurements presented above.

\Cref{tab:varying-participation-levels} reports epoch times as we vary the OP, LP, and JN counts, and two trends confirm \Cref{obs:1}. First, as the population grows, contention-blind methods inflate sharply: ARES over-offloads because it ignores server load, and\replaced{ static}{} SplitFed with few client layers pushes most computation to the server. At $50/50/-$ clients, ARES and low-split SplitFed exceed $400$s, whereas \dynsl and \dynsl-P stay near $198$s and $166$s -- up to a $\sim\!3\times$ reduction. Second, \replaced{no static rule}{such methods do not} track\replaced{s}{} the optimum\replaced{ across regimes}{}: SplitFed(3) and ARES are competitive only at small device counts\replaced{ (the single-type rows)}{ (cf. \Cref{fig:motivation_epoch_time})} and degrade as contention rises, so the two \dynsl variants are the only methods that remain at or near the best throughout. \replaced{Crucially, the profiling-free \dynsl matches or beats the profiling-based \dynsl-P in most regimes and is never far behind, paying only a brief cold-start penalty from initializing at $\delta_{\min}(c)$ while requiring no client benchmarking.}{Without any client benchmarking, \dynsl beats or stays within 4\% of \dynsl-P in seven of ten configurations, but trails it by 13–26\% in the other three, where the cold start is most prominent.}

\begin{table}
\centering
\footnotesize
\renewcommand{\arraystretch}{0.9} % default 1.0
\setlength{\aboverulesep}{0.3ex}   % booktabs default 0.4ex
\setlength{\belowrulesep}{0.45ex}   % booktabs default 0.65ex
\caption{Simulated epoch time (s) for EfficientNet across varying numbers of OPs, LPs, and JNs. SF stands for SplitFed and value in parentheses indicates the number of client layers.}
\label{tab:varying-participation-levels}
\resizebox{\columnwidth}{!}{\begin{tabular}{lll|ccc|c|cc}
\toprule
OP & LP & JN & SF (3) & SF (5) & SF (7) & ARES & DynSL & DynSL-P \\
\midrule
- & - & 10 & 68.8 & 55.4 & 93.3 & \textbf{51.6} & 53.6 & 55.1 \\
- & 10 & 10 & 104.1 & 83.1 & 121.3 & \textbf{83.0} & 99.7 & 88.2 \\
10 & - & - & \textbf{97.0} & 106.5 & 192.8 & 99.6 & 100.5 & 100.9 \\
10 & 10 & 50 & 337.5 & 228.6 & 199.5 & 229.2 & \textbf{107.2} & 113.9 \\
10 & 10 & 10 & 155.0 & 126.4 & 193.5 & 136.4 & 110.2 & \textbf{108.9} \\
10 & 10 & - & 108.8 & 111.7 & 192.8 & \textbf{100.2} & 110.8 & 107.2 \\
- & 50 & 10 & 298.6 & 217.6 & \textbf{125.1} & 251.2 & 126.9 & 132.2 \\
10 & 50 & - & 287.6 & 216.5 & 196.8 & 256.1 & \textbf{130.3} & 139.9 \\
50 & 10 & - & 285.6 & 218.7 & 195.6 & 267.0 & 197.7 & \textbf{157.3} \\
50 & 50 & - & 505.8 & 350.6 & 207.9 & 448.8 & 198.3 & \textbf{166.0} \\
\bottomrule
\end{tabular}}
\end{table}

% \Cref{tab:varying-participation-levels} reports epoch times as we vary the count of OP, LP, and JN clients. \dynsl-P attains the lowest times in every regime; when it is not the best, it is within standard error of the best (we omit reporting standard deviations in the table due to space constraints). \dynsl most often yields epoch times comparable to those of \dynsl-P, however, they are increased due to the initial phase, in which \dynsl runs in a suboptimal configuration. For instance, in the first row, \dynsl-P has epoch time $91.7$ while \dynsl $107.2$, because in the first stage it runs with three layers, and note that without updating the configuration, the epoch time would be $147.0$. ARES under-allocates layers to clients because it ignores server contention, and static \textsc{SplitFed} with few client-side layers is competitive only at small device counts (the single-OP row); as the population grows, the work pushed to the server inflates its epoch time, whereas \dynsl keeps the server off the critical path---directly confirming \Cref{obs:1}.

% \begin{figure}[t]
%     \centering
%     \includegraphics[width=\linewidth]{images/layer_evolution_mobilenet.pdf}
%     \caption{Allocated layers over an epoch with \dynsl on three OPs. The dotted vertical line indicates the time, when ``\emph{OP w/ bg.}'' starts a background process.}
%     \label{fig:allocated-layer-evolution-with-bg}
%     \Description[]{}
% \end{figure}

\subsection{Convergence with a Single Head}
\label{subsec:convergence-on-testbed}

We now train a model until convergence, which we define as the lack of improvement in test accuracy for more than $12$ consecutive evaluation rounds. Following common practice in the FSL literature~\cite{Gao2022-kv,Dachille2024-yx,Chopra2021-vv,Abuadbba2020-fx,Kim2023-wq,Li2021-zc,Poirot2019-ow,Titcombe2021-jm}, we use the Adam optimizer with a constant learning rate of $0.001$.

\subsubsection{Convergence on CoLExT}
\label{subsubsec:convergence-on-colext}
\begin{figure}[t]
    \centering
    \includegraphics[width=\linewidth]{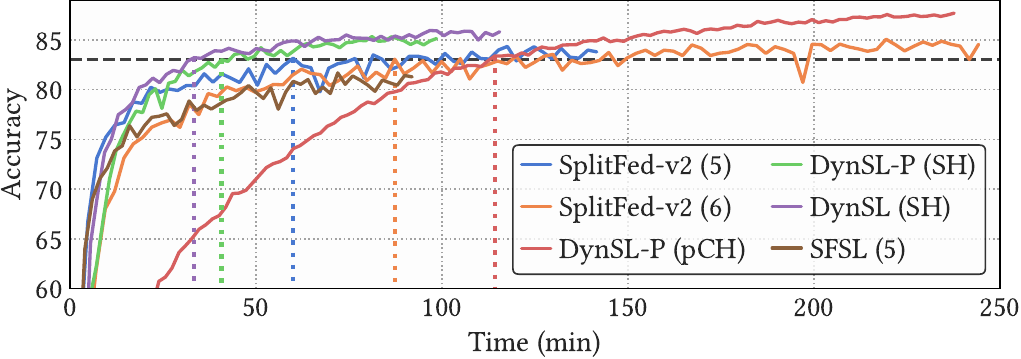}
    \caption{Evolution of accuracy through time when training ResNet18 on CIFAR10. Dashed vertical lines indicate the time-to-accuracy for accuracy $83\%$. For SplitFed-v2 and SFSL, we indicate in parentheses the number of client layers.}
    \Description[]{}
    \label{fig:colext-acc-evol}
\end{figure}

We compare SplitFed-v2, \dynsl, and \dynsl-P with a single\replaced{ classification}{} head\replaced{ on the server}{ (``SH'')}, \dynsl-P with a \replaced{separate client head}{per-client head} (\replaced{denoted \dynsl-P (SH)}{``pCH''}), and SFSL, when training the ResNet18 model. The evolution of test accuracy over time is shown in \Cref{fig:colext-acc-evol}. \replaced{When using a single classification head on the server, }{Among algorithms that train a single head, }\dynsl (SH) and \dynsl-P\replaced{}{ (SH)} accelerate convergence in terms of wall-clock time, which is expected since \dynsl reduces the per-epoch\replaced{ training}{} time as discussed before. For example, \replaced{when targeting an accuracy of $83\%$, }\dynsl\replaced{}{ (SH)} reaches \replaced{this threshold }{$83\%$ accuracy }in \replaced{approximately }{}$\sim\!33$ minutes and \dynsl-P\replaced{}{ (SH)} in \replaced{approximately }{}$\sim\!40$ minutes, while the fastest competing approach (SplitFed-v2 with $5$ client-side layers) requires $60$ minutes.

Interestingly, the nearly $50\%$ reduction in time-to-accuracy over SplitFed-v2 is not explained by shorter epochs alone. Measured per epoch, \dynsl\replaced{}{ (SH)} also reaches the $83\%$ threshold sooner -- in the presented run, at epoch~$17$ versus $24$ for SplitFed-v2~(5). We conjecture that this relates to \dynsl's schedule, under which a larger fraction of the model is trained server-side early in each epoch before layers migrate to clients. We report this as a preliminary observation rather than a general convergence result, and a controlled study is needed to confirm the effect and its cause.

Using \replaced{a separate head per client}{per-client heads} results in slower convergence during the early stages of training\replaced{. However, this configuration}{ but} ultimately achieves the highest final accuracy ($87.6{\scriptscriptstyle \pm 0.3}\%$). The second best accuracy ($85.6{\scriptscriptstyle \pm 0.6}\%$) is attained by \dynsl\replaced{}{ (SH)}, followed by DynSL-P \replaced{}{(SH)} ($85.4{\scriptscriptstyle \pm 0.2}\%$). SplitFed-v2 attains accuracy ranging between $84.1\%$ and $84.9\%$, while SFSL attains $82.0 {\scriptscriptstyle \pm 0.3}\%$ -- we run the experiments three times and report the mean and standard deviation. % \footnote{Standard deviations are obtained by repeating runs $3$ times with different seeds. In \Cref{fig:colext-acc-evol} a single run is shown.} % We repeat such experiments over multiple models and obtain consistent results.

% \begin{figure}
%     \centering
%     \includegraphics[width=\linewidth]{images/colext_acc_evol_efficientnet.pdf}
%     \caption{Evolution of accuracy when training the EfficientNet model on CIFAR10 on CoLExT. Dashed vertical lines indicate the time-to-accuracy for accuracy $83\%$.}
%     \label{fig:efficientnet}
%     \Description[]{}
% \end{figure}

\subsubsection{Accuracy on Different Datasets}
\label{subsec:simulation-accuracy-different-datasets}

\begin{table}
\centering
\footnotesize
\caption{Max. validation accuracy attained by the considered algorithms when training ResNet18 on IID data.}
\label{tab:res-simulation-resnet18-iid}
\renewcommand{\medalsample}{00.0}

\begin{tabular*}{\columnwidth}{@{\extracolsep{\fill}} rr|ccc}
\toprule
 &  & CIFAR10 & CIFAR100 & CINIC10 \\
\midrule
\multirow[c]{2}{*}{DynSL} & One way & \meanstd[2]{87.4}{0.1} & \meanstd[2]{58.3}{0.3} & \meanstd[2]{68.2}{6.0} \\
 & Bidirectional & \meanstd[3]{86.0}{0.2} & \meanstd[3]{56.0}{0.2} & \meanstd[3]{63.6}{8.1} \\
\cline{1-5}
DynSL-P & \replaced{Com. head}{} & \meanstd{78.4}{4.3} & \meanstd{52.5}{0.4} & \meanstd{63.1}{7.6} \\
\cline{1-5}
DTFL &  & \meanstd{72.3}{9.3} & \meanstd{48.1}{0.6} & \meanstd{56.3}{12.0} \\
\cline{1-5}
\multirow[c]{2}{*}{SFSL} & 5 layers & \meanstd{73.4}{7.7} & \meanstd{51.2}{0.5} & \meanstd{55.6}{10.3} \\
 & 6 layers & \meanstd{77.5}{5.5} & \meanstd{51.2}{0.2} & \meanstd{61.2}{10.4} \\
\cline{1-5}
\multirow[c]{3}{*}{SplitFed-v2} & 5 layers & \meanstd{78.3}{4.6} & \meanstd[3]{56.0}{0.1} & \meanstd{62.5}{7.8} \\
 & 6 layers & \meanstd{76.1}{4.1} & \meanstd{50.9}{0.1} & \meanstd{63.3}{7.4} \\
 & 7 layers & \meanstd{74.7}{6.3} & \meanstd{49.6}{1.1} & \meanstd{61.1}{8.8} \\
\cline{1-5}
FedAvg &  & \meanstd[1]{87.7}{0.5} & \meanstd[1]{59.0}{0.2} & \meanstd[1]{78.3}{0.2} \\
\bottomrule
\end{tabular*}
\end{table}

We now simulate training a ResNet18 model across $40$ clients, all of which participate in every epoch. Except for FedAvg, all methods train a single shared classification head on the server. We simulate a training-speed imbalance by having some clients produce embeddings faster than others; consequently, at the start of an epoch the updates from fast clients are more concentrated, while these clients finish sooner.

The goal of this experiment is to take a first, controlled look at how mid-epoch layer migration affects \emph{final accuracy}, rather than the training time studied so far. Because real-world layer movement is stochastic, we here fix a deterministic migration schedule so that all runs are directly comparable. Concretely, for \dynsl we consider two configurations. In \dynsl~``One~way'', representative of the pattern we observe most frequently under normal conditions (see \Cref{fig:layer-evolution-no-bg}), clients progressively receive more layers from the server as the epoch proceeds; this is the favorable regime in which the allocator never over-allocates and no client slows down. In \dynsl ``Bidirectional'', on the other hand, 20\% of the clients decrease the number of trained layers during every epoch. We additionally consider \dynsl-P with a single server-side head.
\begin{table}
\centering
\footnotesize
\caption{Max. validation accuracy attained by the considered algorithms when training ResNet18 on non-IID data.}
\label{tab:res-simulation-resnet18-noniid}
\renewcommand{\medalsample}{00.0}
\begin{tabular*}{\columnwidth}{@{\extracolsep{\fill}} rr|ccc}
\toprule
 &  & CIFAR10 & CIFAR100 & CINIC10 \\
\midrule
\multirow[c]{2}{*}{DynSL} & One way & \meanstd[1]{74.5}{0.5} & \meanstd[1]{50.1}{0.5} & \meanstd[1]{54.4}{1.7} \\
 & Bidirectional & \meanstd[2]{71.3}{0.6} & \meanstd[3]{47.4}{0.7} & \meanstd[2]{54.1}{0.6} \\
\cline{1-5}
DynSL-P & \replaced{Com. head}{} & \meanstd[3]{69.5}{1.9} & \meanstd{45.6}{1.1} & \meanstd{51.1}{1.7} \\
\cline{1-5}
DTFL &  & \meanstd{63.6}{1.7} & \meanstd{37.4}{1.2} & \meanstd{40.8}{3.6} \\
\cline{1-5}
\multirow[c]{2}{*}{SFSL} & 5 layers & \meanstd{58.4}{2.8} & \meanstd{38.1}{1.4} & \meanstd{37.1}{3.5} \\
 & 6 layers & \meanstd{62.9}{1.0} & \meanstd{33.9}{2.0} & \meanstd{39.3}{3.3} \\
\cline{1-5}
\multirow[c]{3}{*}{SplitFed-v2} & 5 layers & \meanstd{66.9}{1.8} & \meanstd{42.4}{0.9} & \meanstd{45.5}{2.5} \\
 & 6 layers & \meanstd{65.7}{2.9} & \meanstd{42.0}{2.4} & \meanstd[3]{53.4}{0.4} \\
 & 7 layers & \meanstd{65.1}{1.3} & \meanstd{46.0}{1.0} & \meanstd{50.4}{0.9} \\
\cline{1-5}
FedAvg &  & \meanstd{59.1}{1.0} & \meanstd[2]{49.3}{0.2} & \meanstd{45.1}{1.1} \\
\bottomrule
\end{tabular*}
\end{table}

We report results for IID data in \Cref{tab:res-simulation-resnet18-iid} and for non-IID data (Dirichlet, $\alpha{=}0.2$) in \Cref{tab:res-simulation-resnet18-noniid}; all numbers are means over five seeds with standard deviations. We read the following trends as \emph{indicative rather than conclusive}. Under IID data, training a separate head per client attains the highest accuracy (reported as FedAvg, and equivalent to the separate-head configuration of \secref{subsec:separate-clf-heads}, which \dynsl can accelerate without altering convergence). Among the single-head methods, \dynsl ``One~way'' is the strongest on all three datasets, essentially matching FedAvg on CIFAR10 ($87.4$ vs.\ $87.7$) and CIFAR100 ($58.3$ vs.\ $59.0$). Under non-IID data the advantage of separate heads no longer holds: FedAvg is not the top performer on any dataset, and on CIFAR10 and CINIC10 it falls among the weakest methods; on CIFAR100 it remains competitive (second best, narrowly behind \dynsl). Across all three datasets, \dynsl~(``One~way'') attains the best -- or, within run-to-run variance, tied-best -- accuracy. Taken together, these runs indicate that \dynsl's normal outward (``One way'') migration does not degrade final accuracy -- never worse than, and sometimes better than, its static single-head counterparts -- while the adversarial \emph{Bidirectional} schedule costs at most $\sim\!5$pp. We stress that these observations are preliminary and that a fuller study\replaced{ across architectures and heterogeneity levels}{} is needed to understand \emph{why} a shared head paired with heterogeneous per-client depths behaves this way. %We regard it as a promising but open direction.

\section{Discussion and Future Work}

\borisparagraph{Convergence with separate heads.} 
With a \emph{separate head per client}, changing the split only redistributes computation, so \dynsl can accelerate virtually any FL algorithm -- by up to $30\%$ in our experiments (Tables \ref{tab:epoch-time-heads}, \ref{tab:time-slowdowns}) -- without altering convergence. The only overhead is transferring the optimizer state on migration to preserve convergence guarantees; empirically this is worthwhile and typically small compared to embedding sizes (e.g., for ResNet18 at batch size $16$, embeddings after the second residual block hold ${\sim}1$M values, versus ${<}700$K parameters in the first two residual blocks).

\borisparagraph{Convergence with a shared head.}
With a \emph{single shared head}, server-side layers update more frequently than client-side ones \cite{Kohankhaki2024-oa,Oh2022-xp}, and the dynamics further depend on the split and client-update ordering, making closed-form analysis hard. Empirically (\secref{subsec:convergence-on-testbed}), heterogeneous per-client depths under a shared head do not harm convergence; a rigorous analysis remains future work.

\replaced{}{\borisparagraph{Alternative cost models.} Metric collection, within-epoch reconfiguration, and sample-aware aggregation are independent of how epoch time is estimated, so our cost model $F$ can be replaced by a richer one if needed. In our setting, however, $F$ proved sufficient: its role is to identify good splits and, through the margin $\xi$, worthwhile reconfigurations, rather than to forecast epoch time exactly. Results throughout \S\ref{sec:evaluation} show it does so: under stable conditions we found no split that consistently outperforms the one \textsc{DynSL}-P selects with $F$, and with the hyperparameters of \S\ref{sec:evaluation} we never observed oscillatory reconfigurations.}

\borisparagraph{Overhead.}
\dynsl introduces three potential sources of overhead, none of which materially affects training time. First, the allocator runs \emph{asynchronously on the server, off the training critical path}, i.e., without stalling forward or backward passes, so its cost is fully overlapped with training; even disregarding this overlap, solving (\ref{eq:optimization_problem}) is cheap (\Cref{claim:poly}), completing in under $0.05$s per invocation across our testbed runs. Second, metric collection piggybacks a handful of scalar timing and bandwidth values onto the embedding/gradient exchange FSL performs anyway, adding no extra round trips and a payload negligible next to the embeddings; reading a wall-clock timer around each pass is orders of magnitude cheaper than the pass itself, and its effect on epoch time falls below the run-to-run variance our devices already exhibit, rendering it indistinguishable from the platform's own noise. Third, reconfiguration may incur the optimizer-state transfer discussed above, which we have shown to be small relative to the data already exchanged.

\borisparagraph{Privacy.}
\dynsl enforces a per-client minimum split as a hard constraint, so no client ever emits embeddings shallower than its privacy floor, \emph{even when a shallower split would be faster}. This floor is orthogonal to, and can be composed with, stronger mechanisms such as differential privacy or noise injection~\cite{Titcombe2021-jm,Abuadbba2020-fx,pham2024enhancing,ohprivacy}.

\borisparagraph{Toward Asynchronous FSL.}
Asynchronous FL suffers from \emph{stale} updates, as slow clients' updates can hurt convergence~\cite{Xu2023-tv,xie2019asynchronous}. Staleness is governed by a client's round-trip time -- the very quantity \dynsl models and controls. This makes the split point a direct lever on staleness: replacing our objective with a per-client deadline gives every client a shallow-enough split to report within a bounded staleness window. Mechanisms similar to \dynsl can thus provide a method to bound stale updates. Adapting \dynsl to this deadline-driven regime is the natural next step.

\borisparagraph{Suitability for Mobile Deployments.}
\dynsl makes no assumption about client compute or network at epoch start, which suits settings where devices continuously join and leave and where only a small fraction of a large population is sampled per round~\cite{Singhal2021-wj}. In such regimes clients participate rarely, so on-device profiling wastes resources it cannot amortize, while allocating work without regard to actual device capacity risks creating stragglers. \dynsl sidesteps both by starting each client at its minimum split and increasing its workload only as measurements arrive, avoiding the profiling cold start while still adapting to device capacity.

\borisparagraph{Limitations.}
In \dynsl, energy is captured only indirectly via the per-client resource ceiling, and setting this budget at runtime is left to future work. A second direction is server synchronization in multi-cloud or hierarchical settings, where disjoint sub-networks of edge devices are each served by a different edge server and the resulting models must be aggregated \emph{across} servers. \replaced{}{We did not evaluate short-lived or oscillating perturbations, nor the sensitivity of \textsc{DynSL} to $\gamma$, $\xi$, and the allocator period, which practitioners may need to tune to their deployment.}
\section{Related Work}
\label{sec:related-work}
\borisparagraph{Adaptive per-Client Split Selection.} The work closest to DynSL selects a per-client split point from measured system conditions. One sub-group fixes the split \emph{once}, before training: CPSL~\cite{Wu2023-bo} and EPSL~\cite{Lin_undated-gr} choose a single global cut layer while jointly allocating wireless resources, and Sun et al.~\cite{sun2025split} select personalized cut layers via a genetic algorithm. A second sub-group re-decides the split \emph{each round}: ARES~\cite{Samikwa2022-fr} and DFL~\cite{Samikwa2024-ia} solve per-round time/energy trade-offs using periodic on-device benchmarking; DTFL~\cite{Mahmoud-Sajjadi-Mohammadabadi2025-pg} re-assigns clients to tiers each round from profiled timing; FedAdapt~\cite{Wu2022-mz} uses a reinforcement-learning agent to pick per-group offloading points; HSFL~\cite{Deng2023-qp} updates a latency model online; AdaptSFL~\cite{Lin2024-ao} jointly optimizes the cut layer and the client-side aggregation interval; and P3SL~\cite{Fan2025-km} lets clients independently pick split points under energy and privacy constraints. In single-client settings, OPEN~\cite{Li2024-xm} applies Lyapunov optimization for layer allocation; OCLA~\cite{Marinova2024-mf} and Zhang et al.~\cite{Zhang2023-ef} select a cut layer from offline profiles, and DSFL~\cite{Wazzeh2025-tr} predicts training time from device metadata. While related, all such methods suffer from the three limitations addressed by \dynsl.%, i.e., they rely on profiling, do not model server contention, and have no mechanism if a client experiences a slowdown.

\borisparagraph{Modifying the Training Procedure.} 
In FSL, some works alter the training objective. FedSplitX~\cite{Shin2023-uy}, SFSL~\cite{Han_undated-pd} and DTFL~\cite{Mahmoud-Sajjadi-Mohammadabadi2025-pg} attach auxiliary networks so clients train on a local loss without server gradients; CFG~\cite{Fu2024-st} clusters clients that train the same number of layers. EPSL~\cite{Lin_undated-gr} averages a fraction of server gradients to shrink back-propagation, and SGLR~\cite{Pal2021-fn} rescales the server learning rate and uses averaged gradients for client backpropagation. PipeSFL~\cite{pipesfl} changes the order in which embeddings are processed by prioritizing embeddings from lagging clients. \dynsl is orthogonal to this line: it leaves the training procedure unchanged, and can in principle be composed with any of these techniques.

% \paragraph{Server-side scheduling and parallelization.} Rather than choosing the split, some systems optimize how the server processes client work. PipeSFL~\cite{pipesfl} prioritizes smashed data from lagging clients; Tirana et al.~\cite{tirana} formulate classification head assignment as an NP-hard ILP; and ParallelSFL~\cite{parallelsfl} groups clients into clusters and tunes per-cluster update frequencies. These levers are complementary to split allocation and could be layered on top of DynSL's allocator.

\borisparagraph{Alternative Collaborative-Training Topologies.}
Pipeline-parallel techniques distribute workload across \replaced{nearby }{}edge devices without involving any server. RingSFL~\cite{Shen2024-yi} replaces the client--server split with a peer-to-peer ring. Related FL works~\cite{liu2022no_inclusivefl,Diao2020-ge,Horvath2021-aa} reduce \emph{model size} rather than offloading computation, assigning sub-models to weaker clients. These approaches are not directly comparable to \dynsl, although they corroborate its premise that heterogeneity is best handled by adapting how much computation each device performs. Lastly, another orthogonal line of research aims to minimize training time by choosing \emph{which} clients participate each round, favoring those with better data utility or system resources~\cite{nishio2019client,lai2021oort}.
\section{Conclusion}

We presented \dynsl, a framework for dynamically distributing computation between clients and the server in collaborative distributed learning settings. By adapting the client–server split based on observed runtime metrics and reacting to changing system conditions, \dynsl reduces the training time of FSL algorithms while preserving -- and in some cases improving -- the final accuracy.

\begin{acks}
\replaced{}{We thank the anonymous reviewers for their constructive feedback, Amândio R. Faustino for his assistance in running experiments on CoLExT, and Saverio Pasqualoni for his help in preparing the figures. This work was in part supported by the Slovenian Research and Innovation Agency through research projects P2-0098 and N2-0393.}
\end{acks}

\balance
\bibliographystyle{ACM-Reference-Format}
\bibliography{cleaned_bibliography}

\end{document}